%% file: 00_main.tex
\documentclass[preprint,journal]{vgtc}       

\ifpdf
  \pdfoutput=1\relax                   
  \usepackage{graphicx}                
  \DeclareGraphicsExtensions{.pdf,.png,.jpg,.jpeg} 
\else
  \usepackage{graphicx}                
  \DeclareGraphicsExtensions{.eps}     
\fi%

\graphicspath{{figures/}{pictures/}{images/}{./}} 

\usepackage{microtype}                 
\PassOptionsToPackage{warn}{textcomp}  
\usepackage{textcomp}                  
\usepackage{mathptmx}                  
\usepackage{times}                     
\usepackage{cite}                      
\usepackage{tabu}                      
\usepackage{booktabs}

\usepackage{paralist}
\usepackage{array}
\usepackage{subfig}
\usepackage{color, soul}
\usepackage[font={scriptsize,sf},tableposition=top]{caption}
\usepackage{float}
\floatstyle{plaintop}
\restylefloat{table}
\usepackage{sparklines}
\usepackage{enumitem}
\usepackage{makecell}
\usepackage{multirow}
\usepackage{multicol}
\usepackage{colortbl}
\usepackage{pifont}
\usepackage{tabularx}
\usepackage{amsmath}
\usepackage[para,online,flushleft]{threeparttable}
\usepackage{cite,etoolbox}

\preprinttext{To appear in IEEE Transactions on Visualization and Computer Graphics.}

\onlineid{1616}
\vgtccategory{Research}
\vgtcpapertype{Representations \& Interaction}

\title{Cultivating Visualization Literacy for Children\\ Through Curiosity and Play}

\author{S. Sandra Bae, Rishi Vanukuru, Ruhan Yang, Peter Gyory, Ran Zhou, Ellen Yi-Luen Do, and Danielle Albers Szafir}

\authorfooter{
\item
 S. Sandra Bae, Rishi Vanukuru, Ruhan Yang, Peter Gyory, Ran Zhou, and Ellen Yi-Luen Do are with CU Boulder. E-mail: \{suba2574, riva3436,ruya6408, pegy8859, razh6791, yido3201\}@colorado.edu.
\item
 Danielle Albers Szafir is with UNC-Chapel Hill. E-mail:  danielle.szafir@cs.unc.edu.
}

\shortauthortitle{Bae \MakeLowercase{\textit{et al.}}: Cultivating Visualization Literacy for Children Through Curiosity and Play}

\def\techname{\textit{Data is Yours}}
\def\childnum{5}
\vgtcinsertpkg

\input{0_abstract}

\teaser{
  \centering
  \includegraphics[width=\linewidth]{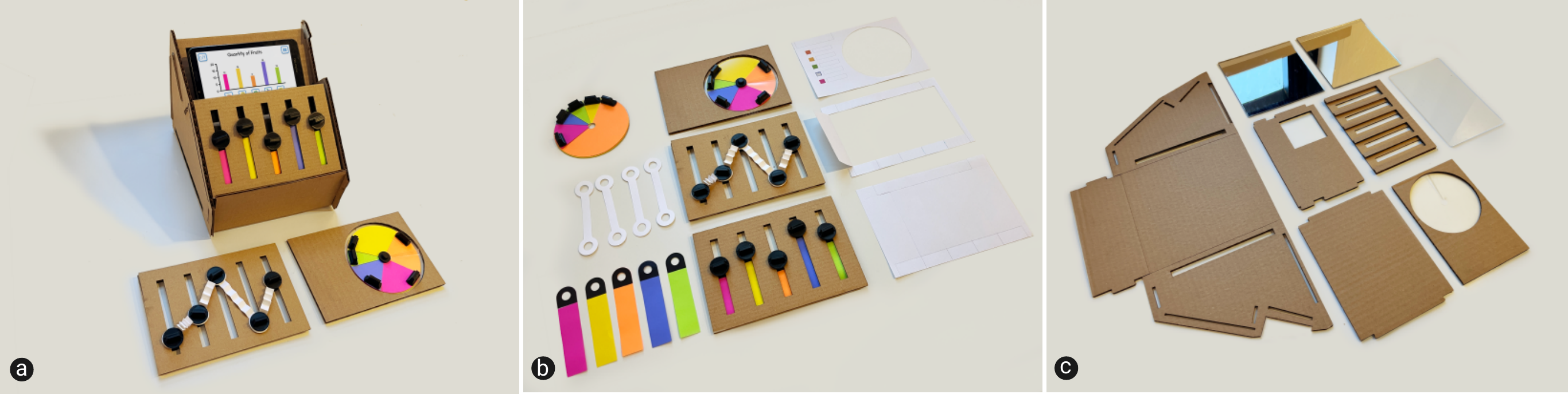}
  \caption{Data is Yours. (a) The assembled toolkit with the physicalization bar chart panel currently in use and a smartphone displaying the rendered visualization. The toolkit supports interchangeable chart panels (i.e., line, pie) that are not in use. (b) Physical components that make up the chart panels and each chart's corresponding paper template. Paper templates for the line and bar charts consist of tear-off rectangles. (c) Physical components that make up the support structure of the toolkit. }
  \label{fig:teaser}
}

\begin{document}
\firstsection{Introduction}
\maketitle

\input{1_introduction}
\input{2_background}
\input{3_formative}
\input{4_application}
\input{5_evaluation}

\input{6_results}

\input{7_discussion}

\input{8_conclusion}

\input{9_acknowledgement}

\bibliographystyle{abbrv-doi}
\bibliography{bib}
\end{document}

%% file: 0_abstract.tex
\abstract{Fostering data visualization literacy (DVL) as part of childhood education could lead to a more data literate society. However, most work in DVL for children relies on a more formal educational context (i.e., a teacher-led approach) that limits children’s engagement with data to classroom-based environments and, consequently, children’s ability to ask questions about and explore data on topics they find personally meaningful. We explore how a curiosity-driven, child-led approach can provide more agency to children when they are authoring data visualizations. This paper explores how informal learning with crafting physicalizations through play and curiosity may foster increased literacy and engagement with data. Employing a constructionist approach, we designed a do-it-yourself toolkit made out of everyday materials (e.g., paper, cardboard, mirrors) that enables children to create, customize, and personalize three different interactive visualizations (bar, line, pie). We used the toolkit as a design probe in a series of in-person workshops with \childnum\ children (6 to 11-year-olds) and interviews with 5 educators. 
Our observations reveal that the toolkit helped children creatively engage and interact with visualizations. Children with prior knowledge of data visualization reported the toolkit serving as more of an authoring tool that they envision using in their daily lives, while children with little to no experience found the toolkit as an engaging introduction to data visualization. 
Our study demonstrates the potential of using the constructionist approach to cultivate children's DVL through curiosity and play.
}

%% file: 1_introduction.tex
\firstsection{..}
Visualizations serve a critical role in public discourse~\cite{claes2015, knudesen2018}, information dissemination~\cite{jones2017data, dork2013critical, zinovyev2010data}, civic engagement~\cite{airinei2010data, bohman2015data}, and decision making~\cite{kokina2017role, murphy2013data}.
The COVID-19 pandemic has emphasized the importance of data visualization literacy (DVL)---the ability to constructively reason with data---as experts and news outlets frequently use visualizations to increase public awareness, communicate public health decision-making, and drive action within different communities~\cite{shneiderman-covid}. The proliferation of visualizations from the pandemic has highlighted how the general public's limited DVL in tandem with poor visualization design (e.g. misleading visualizations~\cite{informatics7030035, doan2021misrepresenting}) is causing a rapid spread of both accurate and inaccurate information \cite{diseases2020covid, 2021-viral-visualizations}.

\vspace{-2pt}
We investigate how physically crafting visualizations (i.e., making and assembling with craft materials) through play and curiosity may serve to make visualization more comfortable and familiar to young children.
Despite significant attention to statistical literacy and, increasingly, general data literacy in elementary curricula \cite{lee2022data}, we still have limited knowledge of how to best foster DVL at early ages. Most work in DVL for children relies on a more constrained, formal educational context (i.e., a teacher-led approach). In addition, we lack formalized methods (i.e., visualization literacy tests for children) to evaluate the efficacy of these approaches. Broadly constraining DVL education to formal environments, children are restricted in where they learn and the datasets they work with (e.g., predefined datasets that children may not generally be interested in). Educational research highlights that allowing children to make artifacts driven by play and their own curiosity (known as a \textit{constructionist approach}) serves as a more approachable entry point for children to better grasp complex topics (e.g., mathematics~\cite{sarama2004building, sarama2004technology, verdine2014deconstructing}, artificial intelligence \cite{kahn2021constructionism}, science \cite{wickham2016constructionism}, and programming \cite{alimisis2009constructionism}). 
Many aspects of DVL develop over prolonged study and experiences, such as the ability to critically reason with and design visualizations. To ground future pedagogical developments, our work explores the constructionist approach to provide preliminary steps towards familiarizing children with visualization.

We investigate \textit{\techname} as a design probe to explore how children can accumulate playful learning experiences and personalize visualizations made from  everyday materials (e.g., paper, cardboard, mirrors) (\autoref{fig:teaser}). \techname\ as a toolkit stems from the constructionist \cite{papert2020mindstorms} and embodied cognition theories \cite{Wilson2002, anderson2012eroding} and lets children lead the learning/creation process. By crafting and playing with physical materials, children can give physical form to data. These physical actions can enable children to consciously reason about a visualization's visual encoding (e.g., colors and marks) freely and without the constraint of finding a ``right'' answer to any guiding task.

\vspace{-1.5pt}
The toolkit's workflow is illustrated in \autoref{fig:workflow}: children assemble the kit, collect data, create a paper template, and then construct a chart panel to represent their data.
Though past methods for fostering children's DVL show promise \cite{alper2017visualization, dasgupta2017scratch, bishop2020construct}, these tools typically rely on exclusively digital solutions. Digital tools may come off as ``black boxes'' to young children and limit their embodied experiences with visualizations \cite{resnick2000beyond}.
In contrast, physical representations enable children to engage in embodied learning and can serve as a more approachable introduction to visualization. Thus, we focus on enabling children to craft physicalizations, where the act of making scaffolds their understanding of different visualization components. 
Our approach maintains the benefits of past digital approaches by using computer vision to dynamically render, update, and save digital versions of the physical charts. 

Using \techname, our objectives are to investigate (1) how children might respond to physical crafting experiences in the context of working with visualizations and (2) how to expand DVL learning to informal learning spaces (i.e., learning that takes place outside of a structured classroom~\cite{oblinger2005leading}).
The toolkit is a product of a three-phase design process. We first conducted a formative evaluation using a preliminary prototype with educators and children to better understand current teaching practices. We then refined the prototype  based on observations and feedback. Lastly, to investigate our two objectives, we deployed the toolkit in a series of in-person workshops with children ($n = \childnum$; aged 6--11) and conducted interviews with educators ($n = 5$).
 
Our observations reveal that the toolkit helped children creatively engage and interact with visualizations. Children with prior knowledge of data visualization reported our toolkit serving as an authoring tool that they envision using in their daily lives, while children with little to no experience found the toolkit as an engaging introduction to data visualization. Educators spoke of the benefits of \techname, emphasizing how it can be easily integrated into existing activities and spaces (e.g., museums, classrooms, homes). From these insights, we provide a future vision outlining how to expand the ways children can creatively engage with the data and visualizations found in their everyday lives.

%% file: 2_background.tex
\section{Background and Related Work}
Our work draws on practices both in education and in visualization. We explore prior approaches and investigate the potential of using constructionist practices to cultivate children's DVL.

\subsection{Tools to Improve Children's Data Visualization Literacy}
\label{sec:toolkit}
\vspace{-2pt}
Focusing on core DVL components\cite{lee2017vlat, boy2014principled, borner2019framework,Chevalier2018visualization}, prior tools for children's DVL help children improve their abilities to interpret and extract information from visually represented data. These tools primarily differ based on their types of activities (i.e., guided vs. child-driven) and intended context (i.e., formal vs. informal instruction).

The majority of existing DVL tools are actively \textit{guided} by an educator and used within a \textit{formal} learning context (i.e., classrooms). For example, C'est La Vis~\cite{alper2017visualization} teaches young children about pictographs and bar charts through concreteness fading---a pedagogical method where new concepts are first presented with concrete examples before progressively abstracting them---via animations. Similarly, Construct-A-Vis~\cite{bishop2020construct} helps children create free-form visualizations within a classroom setting. Both tools consider how teachers can configure datasets and visual encodings for children to work with. While this approach enables teachers to provide guidance and feedback, it largely limits children's engagement with data. Visualizations and data exist in various contexts beyond the classroom (e.g., news, games, books, infomercials, and advertisements). Yet this approach constrains children with which datasets they work with and where they learn.

Other approaches are \textit{child-driven} and used within \textit{informal} contexts (i.e., learning that takes place outside of a structured classroom~\cite{oblinger2005leading}). Gamification approaches, such as Huynh et al.'s role-playing game~\cite{huynh2020rpg} and Diagram Safari~\cite{gabler2019safari}, foster curiosity and experimentation by allowing children to learn through experiential play~\cite{oblinger2004next}. While both games let children pace themselves, they are limited to predefined datasets. In contrast, Scratch Community Blocks~\cite{dasgupta2017scratch} and data sculptures~\cite{bhargava2017data} help children work with datasets that the children think are important. This educational approach stems from the constructionist theory, which advocates for student-centered learning by having children build artifacts that are personally relevant and meaningful~\cite{papert2020mindstorms} (see \autoref{sec:constructionism}).

Like data sculpture~\cite{bhargava2017data} and Scratch Community Blocks \cite{dasgupta2017scratch}, we follow the constructionist approach. However, we investigate this approach from a do-it-yourself (DIY) perspective, drawing on constructive visualizations \cite{huron2014constructiveviz}. Manually constructing visualizations has helped adults explore~\cite{huron2014constructiveviz, huron2014constructingtokens}, learn~\cite{perin2021students, vandemoere2010physical}, reflect~\cite{gourlet2017cairn, thudt2018reflect}, and play~\cite{hopkins2020craft} with data. Children may similarly benefit from this approach. By constructing an artifact, children have an ``object-to-think-with'' to externalize their mental model, observe when their designs succeed and fail, and discuss their insights with others~\cite{papert1993children}. Research also shows that while modern technology provides rich functionalities, it may come off as ``black boxes'' to young children as its inner mechanisms are often hidden and thus poorly understood~\cite{resnick2000beyond}. To circumvent this challenge, like data sculpture~\cite{bhargava2017data}, we rely on low-cost physical materials that children can manipulate to create interactive physicalizations and learn from their embodied experiences.

Working with physical materials leverages the embodied cognition theory, which states that motor action and cognition are highly interrelated, and hence, mutually dependent upon each other~\cite{Wilson2002, anderson2012eroding}. 
However, working with physical materials can be challenging and tedious when needing to update datasets and perform other data interactions (e.g., filter, select, encode). As a result, physicalizations traditionally limit children to represent mainly static datasets \cite{Jonathan2020teaching, Verhaert2021datablokken}. Our approach overcomes this limitation and differs from data sculptures in that we still leverage computational capabilities where children can dynamically render, update, and save their visualizations (see \autoref{sec:digital}).

\subsection{Using Toolkits to Construct Visualizations}
\label{sec:constructionism}
Papert postulated that making and designing are critical to learning: they position the learner as an active agent in the creation process, rather than as a passive recipient of materials designed for the learner~\cite{papert2020mindstorms}. D'Ignazio \& Bhargava~\cite{d2016databasic, d2018creative} recognized the benefits of the constructionist model when teaching visualizations to novices. They focused on hands-on learning with creative activities (e.g., analyzing music lyrics) to introduce new data concepts.
However, their work explicitly targets adults. 
Research with children requires a different set of considerations \cite{punch2002children, druin2002role}, making D'Ignazio \& Bhargava's activities difficult to apply to children. Alternative considerations are needed to create meaningful artifacts for children's diverse interests and experiences. 

Toolkits can help achieve the twin goals of meeting children's idiosyncratic needs and constructing visualizations in a creative, playful manner. Learning technologies illustrate that creativity support toolkits can support a wide range of explorations powered by children's interests~\cite{resnick2005design}. The success of these toolkits is driven by design principles, such as being approachable and supporting exploration, where children can construct diverse artifacts even with low-cost, easily available materials (e.g., Cricket~\cite{resnick2000beyond}, Makey Makey~\cite{silver2012makey}). Visualization also highlights the benefits of toolkits, such as the InfoVis Toolkit~\cite{fekete2004toolkit}, prefuse~\cite{heer2005prefuse}, and D3~\cite{bostockd3}. They can simplify creating complex visualizations using methods that facilitate critical thinking and ownership \cite{mendez2017bottom}.
Creative use of visual toolkits, such as danceON~\cite{payne2020dance}, shows how creative activities can engage novices (e.g., young women of color) in technical computational and data concepts while still providing personal and culturally relevant learning experiences. \techname\ has similar objectives. We explore how to best support children to playfully and creatively construct visualizations while broadening their engagement with data through physical construction.

We investigate the potential of physically constructing visualizations for DVL as it can help children become more comfortable and familiar with visualizations.
By learning from embodied experiences (see \autoref{sec:toolkit}), children may understand the different components of a variety of visualizations.
In turn, this approach may help children read and interpret existing visualizations. However, toolkits to help children author visualizations driven by their interests are limited and sparse (\autoref{sec:toolkit}). We take inspiration from both education and visualization to create a toolkit using everyday objects (e.g., cardboard, paper, mirrors) and paper templates to empower children to create, customize, and personalize their own visualizations. Grounded by constructionist and embodied cognition theories, we use \techname\ to explore how to leverage child-driven activities to author interactive visualizations and expand DVL learning to informal learning spaces. Our approach offers a new perspective on how to best cultivate children's DVL through creativity and play.

%% file: 3_formative.tex
\begin{figure}[tb]
	\centering
	\captionsetup{farskip=0pt}
	\includegraphics[width=\columnwidth]{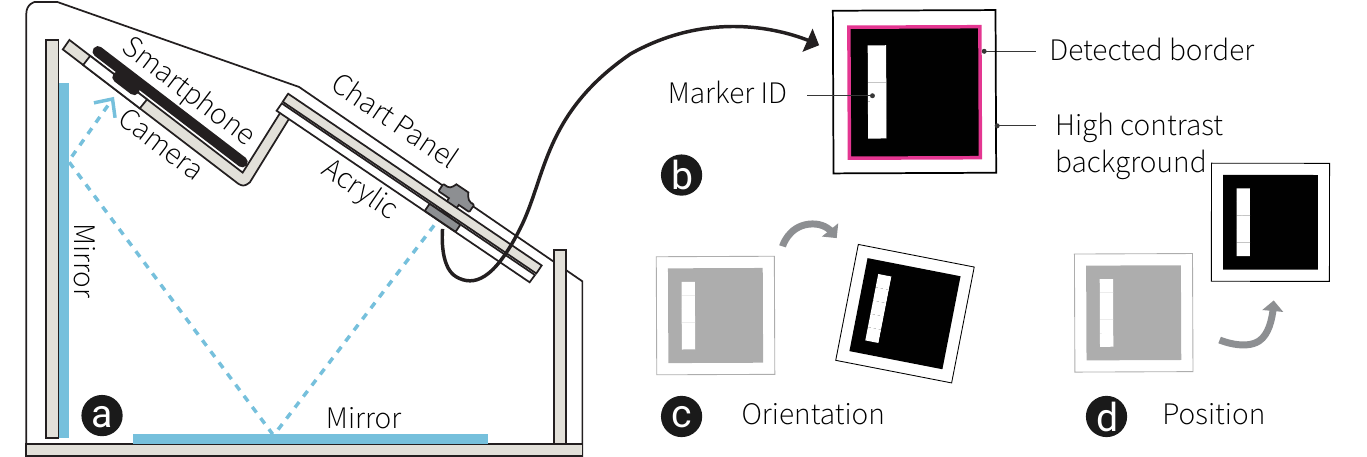}
    \caption{(a) Diagram of inner workings of \techname. (b) Fiducial markers (FM) behind chart panels face the phone's rear-facing camera. Detected FM (outlined in pink) require a high-contrast background. FM are detected based on their (c) orientation and (d) position.}
    \vspace{-0.2cm}
	\label{fig:markers}
\end{figure}

\vspace{-2pt}
\section{Formative Study}
We developed an initial prototype and used it to conduct a formative study with children and educators. The prototype was designed to reach a broad audience using prevalent technology (e.g., smartphones). The objectives of the formative study were to (1) understand the methods and challenges associated with teaching data visualizations in elementary school and (2) collect feedback on the prototype to inform the development of a design probe for an exploratory study.

\vspace{-2pt}
\subsection{Formative prototype}
Inspired by the various tangible interactions printed fiducial markers enable~\cite{zheng2020paper, electronics10050604}, our preliminary prototype explored these markers as control points for children to map, read, and update paper-based visualizations~\cite{bae2021idc}. Each fiducial marker provides information based on its identity, position, and orientation (\autoref{fig:markers}).
Our initial setup included a single USB camera mounted above the paper graph. The camera was connected to a laptop which presented the digital visualizations.

\begin{figure}[tb]
	\centering
	\captionsetup{farskip=0pt}
	\includegraphics[width=\columnwidth]{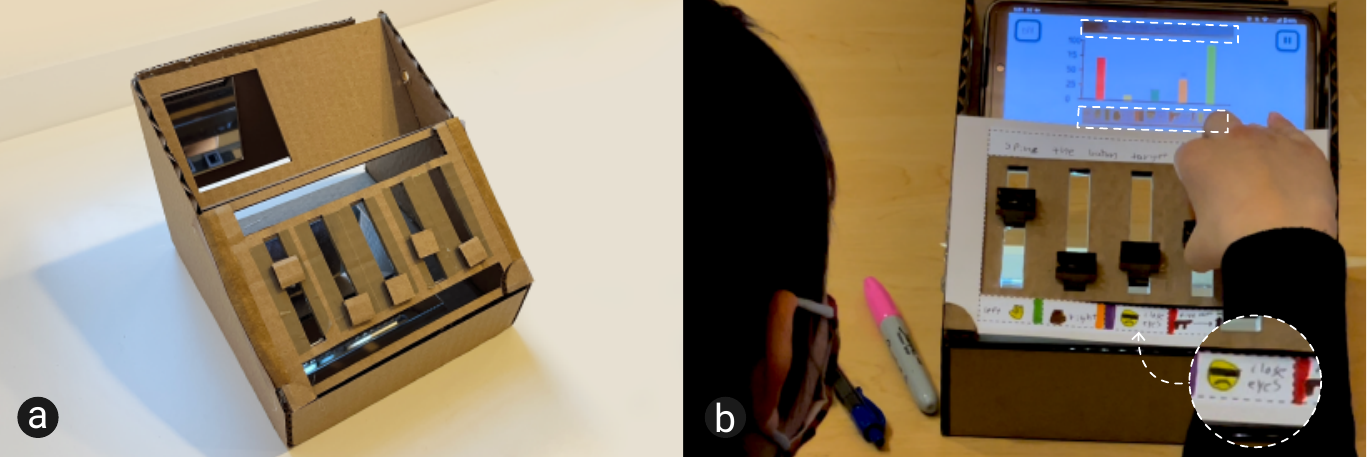}
    \caption{(a) Formative prototype. (b) A child from formative study is moving cardboard sliders to author a bar chart. His paper template consists a mix of words and icons and is reflected on the smartphone display (indicated by dotted outlines). See \autoref{sec:flow2} for more details.}
    \vspace{-0.2cm}
	\label{fig:formative}
\end{figure}

However, to enable participation from a broader set of backgrounds, we designed a new prototype that used a smartphone. The smartphone acted both as a camera to track changes in markers as well as an interactive screen to show a digital representation of the physical visualization. Relative to personal computers, smartphones are considered to be more widely accessible hardware---in 2019, 95\% of U.S children had internet access at home, but 6\% (primarily low-income minorities) accessed the internet only via smartphones~\cite{nces2021}. 
The prototype consisted of five vertical sliders that were mapped to five bars of a bar chart (\autoref{fig:formative}a). 
Fiducial markers were placed behind each slider, and the smartphone was mounted at an angle that can view the fiducial markers via two mirrors enclosed in the cardboard casing (\autoref{fig:markers}a). 
To further enhance authoring capabilities, we introduced paper templates that children could draw or write custom chart titles and labels. The templates would first be scanned by the phone camera and then applied to the digital chart (\autoref{fig:formative}b).

\vspace{-2pt}
\subsection{Method}
\label{sec:formative_methods}
After designing a phone-based prototype, we conducted a formative study with children and educators, which was approved by CU Boulder IRB.
All study materials are available in the supplemental materials\footnote{\href{https://osf.io/7u93z/}{https://osf.io/7u93z/}}.

\textbf{Children}. We interviewed two children (ages 10 \& 11; 1 female, 1 male). We first conducted a pre-interview using images of visualizations from popular media to determine their familiarity with visualizations and asked how the children currently create visualizations. We then used a think-aloud protocol asking participants to work through two activities for approximately 30 minutes. For each activity, we asked what data observations children noticed from the charts they created. In the end, we conducted a semi-structured interview to understand the children's experience in creating their own visualizations and learn how the prototype could be improved. We took notes and captured images and video recordings for analysis.

The first activity taught the children how to use the prototype. Children first sorted different cards with fruit images. Then the children entered the number of each fruit type to create a bar chart using the prototype's sliders. In the second activity, children generated their own data from a game. Children were given toy disks to slide toward a target (see \autoref{fig:workflow}b). The closer the disk was to the center, the more points children were rewarded. The game was played under five different conditions. Four of the conditions were predefined by the research team (left hand, right hand, closed eyes, five steps back), and one was child-defined (i.e., the child created their own game condition). Children created a bar chart of the points they scored per condition using the prototype and the paper templates.

\textbf{Educators}. To complement our formative study with the children, we interviewed 5 educators (3 female, 2 male) who work in a variety of settings: educational research organizations, formal classrooms, after-school programs, and online teaching. The objectives with the educators were to understand current practices, the challenges they faced when teaching data visualization, and how the prototype might help address those challenges. We conducted a semi-structured interview and an artifact walk-through where we shared the same prototype the children used and demonstrated the two aforementioned activities. We took notes and captured images and video recordings for analysis.

\vspace{-2pt}
\subsection{Key Findings and Design Implications}
We identified three themes from the formative study: a need for approachability, engagement, and scaffolding and embodied learning.

\textbf{Approachability}. 
Prior studies and our conversations with educators note that children even of the same general age range are likely to have a variety of experiences with data, drawn from both their personal experiences and educational backgrounds. This variety means that there is no standard method to teach children about data visualization~\cite{lee2022data}. 

Though it is difficult to determine how to make visualization approachable for all children, we noted the promise of everyday materials.
Educators commented that because the prototype was made from materials that children are familiar with (e.g., paper), it can readily foster experimentation (i.e., if a child tried something and did not like the results or made a mistake, they could simply start over). We observed this during the study where children freely engaged with the prototype without fear of breaking something.

\textbf{Engagement}. Educators and children alike expressed a need to center the child’s learning around their interests. This is particularly pertinent as young children have short attention spans, yet for them to learn, they need to be driven by frequent, long-term engagement. Educators address this challenge by empowering students with autonomy and increasing their sense of ownership (e.g., letting children choose their own project topics). We saw this reflected in the formative investigation where both children found the first activity with the predefined fruit dataset as ``cheesy'' and one they would not be interested in doing again. Educators also suggested letting children construct parts of the prototype themselves to increase their investment in the learning activities.
From these comments, the final design probe expanded children’s ability to author their own visualizations in two ways. First, we designed the toolkit such that children can freely author visualizations using data of their choice~\cite{pasek2015learners}.
Second, we designed the probe for easy assembly and provided more options to personalize each chart (e.g., choosing colors and the number of data points) to increase their sense of ownership in the activities.

\textbf{Scaffolding \& Embodied Learning}. Educators highlighted how students struggle when initially designing visualizations (e.g., determining what information should go on which axis, setting up appropriate scales), remembering visualizations components (e.g., titles, axes labels), and interpreting visual representations (e.g., reading values between y-axis tick marks). These remarks generally aligned with our observations that children \textit{recognized} common visualizations (bar, line, pie) but sometimes struggled to articulate what they were conveying.
Educators also noted that though the prototype's input method of connecting physical movements to abstract changes in the bar's height supports embodied learning (see \autoref{sec:toolkit}), more explicit associations between the physical form and visualization would be useful.

Based on these comments, we designed the probe to tightly integrate contextual scaffolding into the physical artifact. We changed the design from a pre-assembled tangible interface to a DIY toolkit where children would physically construct the box and charts to create physicalizations. The chart panels and paper templates of the design probe were designed to be a complete visual representation on their own by containing all necessary information (e.g., visual encoding, title, axes) while still corresponding to the digital representation of the visualization. 
Scaffolding is built into the design of the physical charts that children would reflect on while constructing (e.g., choosing the number of sliders as a proxy for defining the number of data points; \autoref{sec:physical-components}).

\vspace{-2pt}
\subsection{Design Goals}
\label{sec:design-goals}
We derived three core principles from our formative investigation and prior literature: the design probe should (1) be approachable to children; (2) provide children an engaging experience when authoring visualizations
and (3) support scaffolding and embodied learning. 
We further divide these principles into six design goals to understand the usability, inclusiveness, and widespread applicability of the design probe in our exploratory study.\vspace{7pt}

\vspace{-1pt}
\begin{compactdesc}
  \item[\textnormal{\textit{Approachability}}]
  \item[\textbf{G1.}] \textbf{Foster experimentation:} Use low-cost materials to encourage children to experiment and iterate on designs without having to worry about losing or damaging materials.
  \item[\textbf{G2.}] \textbf{Replicable:} Create a design with materials that can be replicated at home or informal learning spaces (e.g., after-school programs, libraries) to expand data engagement.\vspace{7pt}
\vspace{-1pt}
 \item[\textnormal{\textit{Engagement}}]
  \item[\textbf{G3.}] \textbf{Ease of use:} Create a low barrier of entry for children to build the toolkit and explore a wide range of datasets.
  \item[\textbf{G4.}] \textbf{Support creative expression}: Enable children to create their own designs when authoring visualizations.\vspace{7pt}
\vspace{-1pt}
  \item[\textnormal{\textit{Scaffolding \& Embodied Learning}}]
  \item[\textbf{G5.}]  \textbf{Craft visualizations:} Provide a contextualized, embodied experience to help children understand different visualization components.
  \item[\textbf{G6.}] \textbf{Physical interactions:} Focus on making kit mechanisms transparent, connect physical manipulations to data operations, and help children understand why different features are necessary.\vspace{5pt}
\end{compactdesc}
\vspace{-1pt}

%% file: 4_application.tex
\vspace{-4pt}
\section{Design Probe: Constructing Visualizations with Everyday Materials}
To address these goals, we created \techname: a design probe to study how constructionist practices can engage children to create visualizations in a playful, creative manner. It is a toolkit made out of everyday materials (e.g., paper, cardboard) and uses fiducial markers to help children author visualizations. \autoref{fig:workflow} illustrates the workflow of using the toolkit. After constructing the kit (G6), children mount a smartphone on top of the box and place a chart panel with a paper template onto the acrylic surface. Then children manipulate the physicalization to reflect their collected data and are able to see a corresponding visualization on the smartphone display. The toolkit is open source\footnote{\href{https://www.instructables.com/Data-Is-Yours-Toolkit-Assembly-Instructions/}{https://www.instructables.com/Data-Is-Yours-Toolkit-Assembly-Instructions/}} to enable children, parents, and educators to adopt and fabricate the toolkit on their own (G2).

\subsection{Physical Components}
\label{sec:physical-components}
The physical components of the toolkit include  interchangeable chart panels, paper templates, and a cardboard-based box. These components are made from readily accessible materials that can be easily replicated  (G2, G3). 
Children first decide the number of data points required, assemble the chart panel, and then create a paper template to label the chart. The combination of the chart panel and the paper template results in a complete physicalization (\autoref{fig:chart-panels}).

\vspace{-2pt}
\subsubsection{Chart Panels}
\begin{figure}[tb]
	\centering
	\captionsetup{farskip=0pt}
	\includegraphics[width=\columnwidth]{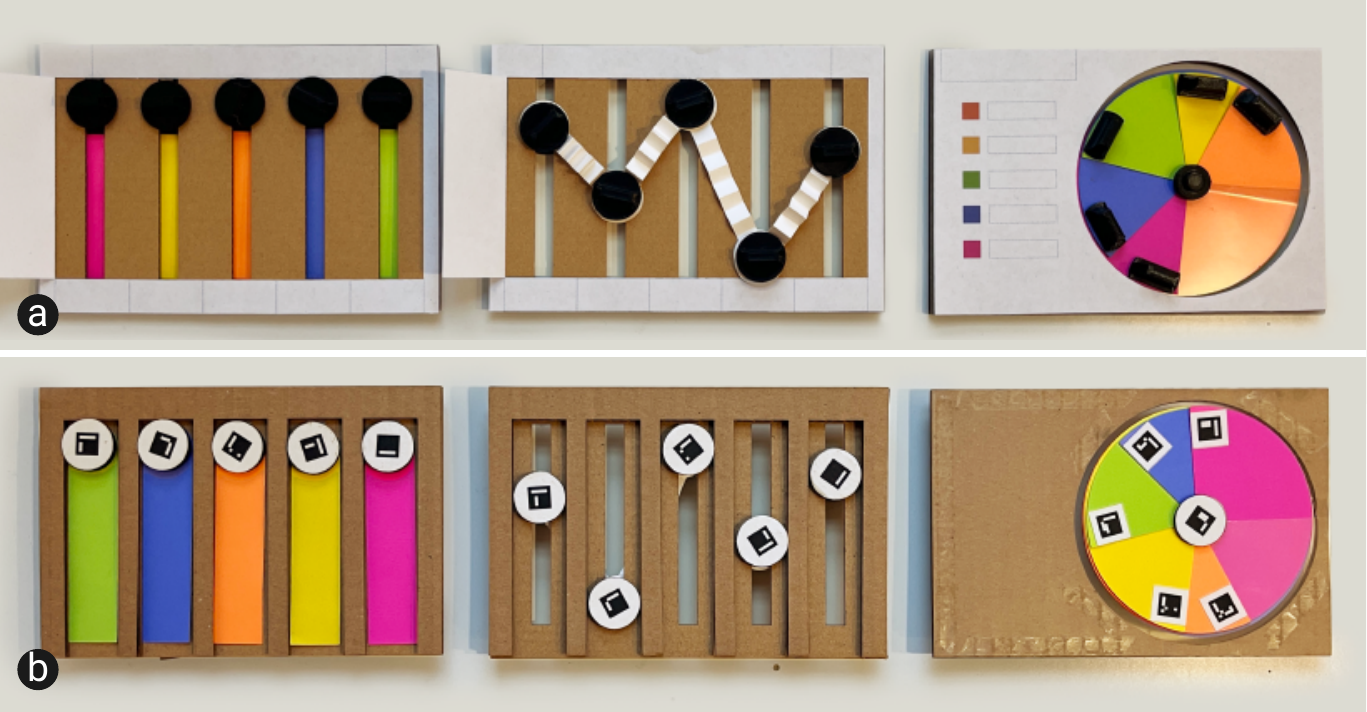}
    \caption{(a) Frontside of each chart panel (bar, line, pie). Blank paper templates layered on top. (b) Backside of each chart panel, illustrating where fiducial markers are placed when facing the camera (\autoref{fig:markers}). }
    \vspace{-0.2cm}
	\label{fig:chart-panels}
\end{figure}

The design probe supports three chart types (bar, line, pie) (\autoref{fig:chart-panels}) (G5). These charts were chosen based on their common use in media and our insights from the formative study. Each chart panel consists of a cardboard base, paper, and 3D printed sliders. To support scaffolding and embodied learning, children use these physical pieces to assemble a visualization. Children can draw connections of the existing idioms behind common visualizations through their physical interactions as opposed to being presented with a predefined chart. Note that though \autoref{fig:chart-panels} shows five data points for each chart, charts can be customized so that children can remove and add data points accordingly. A paper-based alternative to the 3D printed components is provided in our Instructable documentation.

\textbf{Bar chart.} The bar chart panel uses two cardboard panels, which are held together by 3D printed snap presses. Strips of colored paper are inserted between the cardboard and the handle base, acting as the physical ``bars'' of the bar chart. To support creative expression and help children think about visual attributes, children can choose different color strips of paper while authoring a bar chart (G4). For example, children can reason about visual semantics when five bars each have a unique color as opposed to the same color. Fiducial markers are attached to the back of the 3D printed snaps to track the height of the bars while children are physically manipulating them.

\textbf{Line chart}. The line chart panel uses the same cardboard body and 3D printed handles as the bar chart panel. This approach allows children to think about the relationship between a bar and line chart (i.e., how a bar chart can be transformed into a line chart with a temporal emphasis) through hands-on assembly (G5). Instead of wide paper strips, the line chart uses narrow paper strips that are placed between the cardboard and the 3D handle. These paper strips have pre-punched folding lines that shrink and stretch as the 3D handles move either closer or further away from each other. The paper strips are intentionally designed to help children observe and experiment with the strips as they move the 3D printed handles (G1).

\textbf{Pie chart.} The pie chart panel requires a cardboard base and a transparent plastic sheet. Children assemble the “slices” of a pie chart using cardstock circles. Each cardstock circle corresponds to a section of the pie chart and has a unique fiducial marker at the back. These cardstock circles are spirally inserted into the slit of the transparent plastic sheet and pinned by a 3D printed clip. Children can then rotate and manipulate the pie chart around the 3D printed clip. Children can choose pie sections of any color and arrange them in order of their preference.

\subsubsection{Paper Templates}
\label{sec:paper-templates}
Each chart panel requires a paper template, which is a layer of paper that covers the top of the chart panel (\autoref{fig:chart-panels}a). The paper templates are made from regular printing paper (G2), and children can be readily provided with multiple copies of the same paper templates to foster experimentation (G1). The paper templates' design subtly scaffolds children about visualization design by having them fill in the slots for different chart components (e.g., titles, axes labels, y-axis scale). The blank templates also enable children to creatively customize the charts (e.g., by drawing, writing) (G4).

The paper template design for the bar chart and line chart is the same. It consists of a title block, five x-axis label blocks, and tear-off rectangles. The left-most rectangle folds back so children can label the y-axis scale. The tear-off rectangles are to encourage children to think about the data they are representing (G5). Before creating the physicalization and rendering the visualization, children must decide how many rectangles to tear off depending on the number of their data points. Children can tear off up to four rectangles, and those that are not torn will cover the sliders of the respective chart panel. The low cost of paper allows children to experiment and try again if they make the mistake of ripping off too many rectangles (G1).

The paper template for the pie chart similarly includes a title block but features a color legend and a cut-out circle. The color legend helps children consider what the different colors represent within the pie chart. Children can also take notes on their pie chart in the white space of the template. For unused colors, we provided white stickers for children to cover the corresponding legend blocks.

\subsubsection{Box}
The main support structure of the toolkit is a folded cardboard box (\autoref{fig:teaser}c). Four other cardboard components are required to create the final structure. These five components can be quickly assembled and disassembled by simply inserting and unfastening tabs without needing glue or other adhesive materials. The box was designed so that the assembly does not require fine motor skills, making it more generally approachable and appropriate for children (G3). In addition, the cardboard material allows children to decorate the box, promoting creative expression while providing a sense of ownership (G4). 

Inside the assembled box, children place two mirrors to capture the fiducial markers with the phone camera (\autoref{fig:markers}a). Afterward, children insert an acrylic plate from either side of the box. The acrylic panel acts as a support for the interchangeable chart panels. Then the chart panels and paper templates are directly placed on top of the acrylic plate. 

\subsection{Digital Components}
\label{sec:digital}
\begin{figure}[tb]
	\centering
	\captionsetup{farskip=0pt}
	\includegraphics[width=\linewidth]{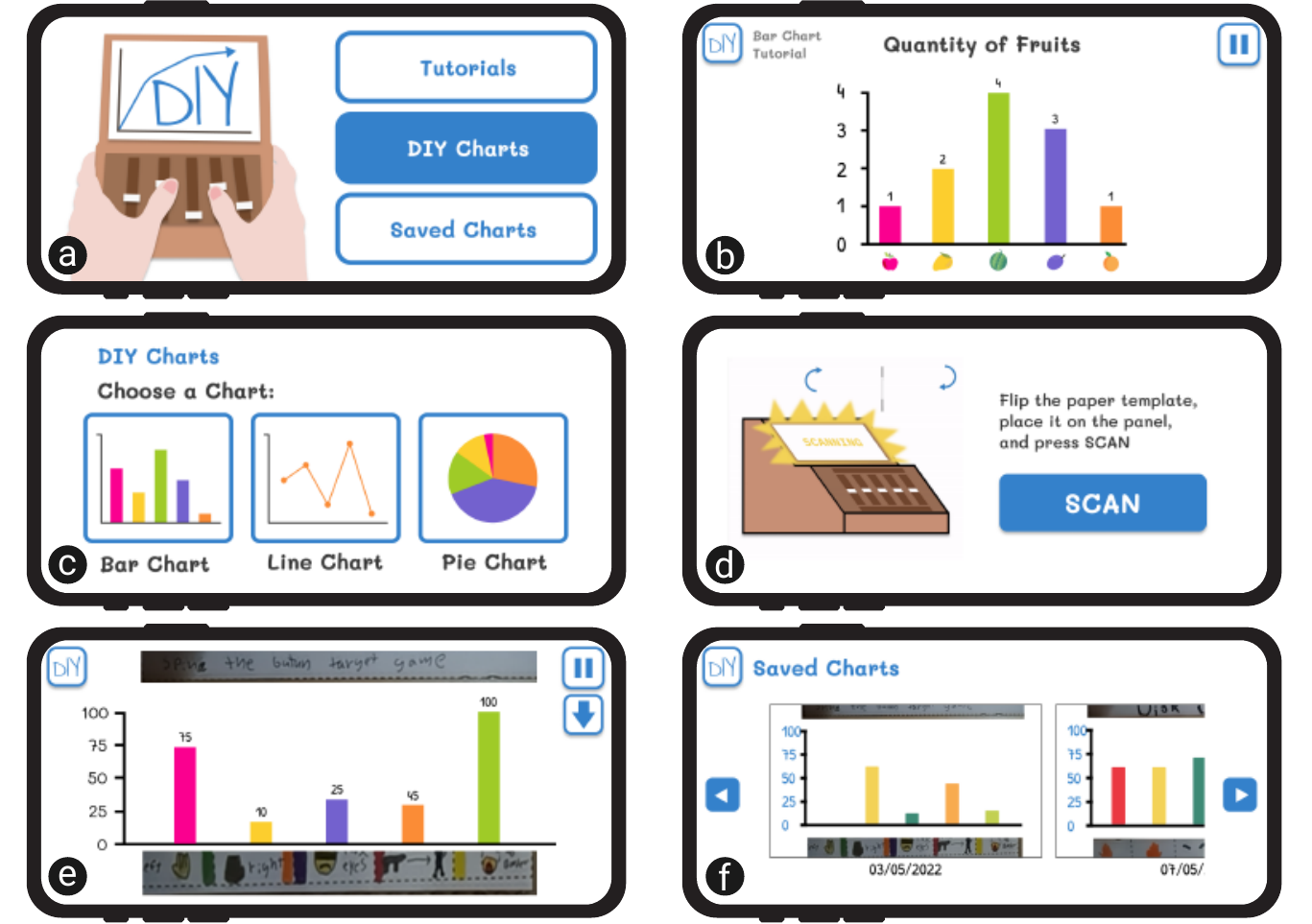}
    \caption{Screenshots of web app. (a) Home page (b) Bar chart tutorial in Flow 1 (c) Menu in Flow 2 (d) Scanning instructions for paper templates (e) Example of a scanned template (f) Saved visualizations in Flow 3.}
    \vspace{-0.2cm}
	\label{fig:app}
\end{figure}

\techname\ supports a web app where children can dynamically render, update, and save their visualizations. The app consists of three user flows (\autoref{fig:app}) that children enter after constructing the toolkit and creating their physicalizations.
The first flow teaches children how to interact with the chart panels to author a visualization, the second flow enables children to scan and save their own chart, and the third flow allows children to revisit previously saved charts. All three flows start via their respective buttons on the home screen. 

\subsubsection{Flow 1: Interacting with the Chart Panels.} Flow 1 provides a tutorial of the system's digital functions and represents the first experience children have when connecting the physicalization to the digital visualization. After constructing and placing the physical chart panel on top of the acrylic panel, children can explore the basic features of their respective digital charts. 
Children can move the physical components and see how the digital representations change as a result. They can also change the color of the digital chart by tapping on the visualization components, such as an individual bar, to match the chosen colors represented in the physical panel.

Children can also ``pause'' the chart in its current position and lift the phone to share their rendered visualizations. The pause functionality prevents the digital chart values from incorrectly changing as the phone camera's field of view is moved around. Given Flow 1's purpose as a tutorial, the data types and scales in Flow 1 are the same as those in the formative study: visualizing different types and quantities of fruit with a predefined dataset. 

\subsubsection{Flow 2: Authoring Visualizations}
\label{sec:flow2}
Once children have created the physical panels, collected relevant data of their choice, and prepared the corresponding paper templates, they can enter Flow 2 to author interactive visualizations. Flow 2 requires two steps: scanning and authoring.

\textit{Scanning and defining the chart.} Children are guided on how to scan their paper template through a series of GIFs and instructional text displayed on the screen. To scan their paper templates, children remove the physical panel and place the paper template face-down on the acrylic panel. The app displays a ``Scan'' button to initiate the process. Once the scanning is complete (i.e., taking a picture), children are instructed to flip the paper template upright, reposition the physical panel, and place the template on top of the chart panel to complete the physical chart. For only the line and bar chart, children are prompted to enter information corresponding to the x- and y-axes of the charts, such as the number of data points  and the maximum value of their collected dataset.

\textit{Authoring and interacting with the chart.} To author visualizations, children follow the same general process as Flow 1. The predefined dataset is replaced with information from the paper template that the children scanned into the chart. Information from the templates, such as the hand-written chart title and x-axis labels, are displayed at the corresponding locations on the screen by taking cropped images. The scale of the y-axis is calculated based on the input from the previous step. Children can then move the physical sliders to represent the data values they have collected. 
This information is translated onto the digital visualization.
In addition to the ``Pause'' button, Flow 2 also allows children to save a chart. Clicking the ``Save'' button results in a small visual indicator that the chart in its current state has been saved.

\subsubsection{Flow 3: Revisiting Saved Charts} Once children have saved their charts, they can view them within the web app by selecting the ``Saved Charts'' button from the home screen. They can browse through a list of chart images in order of the day and time at which they were saved. These images are non-interactive and meant to serve as ``snapshots'' of data collected in time.

\subsection{System Architecture \& Implementation}
Once the toolkit is fully constructed, we have two mirrors inside which direct the phone's rear-facing camera at the chart panels (\autoref{fig:markers}a). This configuration enables a web-based application to detect the markers on the interface and render the charts. This approach allows us to make a variety of interfaces that function with just paper. The front-end is implemented with a combination of HTML5, CSS, JavaScript, and a Javascript port~\cite{js-port} of the ArUco Computer Vision Marker library~\cite{garrido2016generation, romero2018speeded}. The library offers up to 1000 unique markers for tracking a variety of values (e.g., position, rotation).

%% file: 5_evaluation.tex
\begin{figure*}[tb]
	\centering
	\captionsetup{farskip=0pt}
	\includegraphics[width=\linewidth]{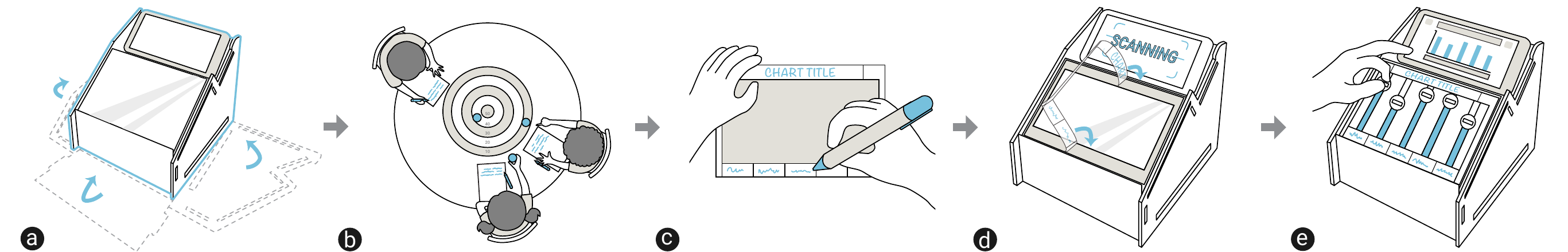}
    \caption{Workflow for children in the formative and exploratory study. (a) Construct a kit (b) Collect data by playing the game of disk darts (c) Author a paper template (d) Scan the paper template (d) Manipulate the physicalization chart panel (bar, line, pie) to match collected data.}
    \vspace{-0.2cm}
	\label{fig:workflow}
\end{figure*}

\section{Exploratory Study}
To investigate how constructionist practices can foster children's comfort and familiarity with data visualization, we conducted an exploratory study using the design probe in a series of workshops with children (\autoref{sec:kids_eval}) and interviews with educators (\autoref{sec:teachers_eval}). Workshops can elicit rich qualitative insights for early stages of applied visualization research~\cite{kerzner2019workshops}. Using guidelines by Kerzner et al.~\cite{kerzner2019workshops}, Huron et al.~\cite{huron2016using, huron2017workshop}, and Terre des hommes~\cite{terre2020focus}, we designed our workshop around the constraints of working with children in a limited timeframe while still evaluating the research objectives of the toolkit. 
Note: the workshop is a means of investigating the constructionist approach where we primarily collected qualitative data rather than quantitatively testing children’s learning outcomes. Current DVL tests are meant for adults and their efficacy for children has not been fully researched\footnote{Existing standardized tools\cite{boy2014principled, lee2017vlat, borner2019framework} target adults. E.g., two components of VLAT focus on determining range and extremum for various visualizations. Such concepts are generally not taught till children are older (11--14 y/o).}.

All activities with the children and educators were approved by the CU Boulder IRB. Study materials are available in supplemental materials. A comprehensive overview of the participants can be found in \autoref{tab:participants}.

\begin{table}[tb]
\small
\begin{tabular}{
                >{\centering}
                p{2cm}
                >{\raggedright}
                p{1.5cm}p{4cm}
                }
    \toprule 
 \textbf{Participant} & \textbf{Role} & \textbf{Age / Occupation}\\\midrule

    P1 &
    Child &
    6 years old\\
    
    P2 &
    Child &
    6 years old\\
    
    P3 &
    Child &
    7 years old\\
    
    P4 &
    Child &
    8 years old\\
    
    P5 &
    Child &
    11 years old\\
    
    E1 &
    Educator &
    After-school program\\
    
    E2 &
    Educator &
    After-school program\\
    
    E3 &
    Educator &
    After-school program\\
    
    E4 &
    Educator &
    Educational Researcher\\
    
    E5 &
    Educator &
    Elementary School Librarian\\
   
\bottomrule
\end{tabular}
\vspace{-0.5em}
\caption{Summary of participants' backgrounds in the exploratory study.}
\label{tab:participants}
\vspace{-0.1cm}
\end{table}

\subsection{Workshops with Children}
\label{sec:kids_eval}
\textbf{Participants}. We conducted two two-hour workshops with children ($n = 5$) between 6--11 years old to understand how the constructionist approach may differently impact different age groups. This age range aligns with the development stage where children are able to formulate representational thought through symbolic thinking (i.e., words and pictures) \cite{goswami2002blackwell}.
The first workshop was a group workshop at an after-school program with four children (2 male, 2 female; aged 6--8 years old), and the second workshop involved one child (1 female; aged 11 years) at a public museum. 
We recruited the participants by sharing virtual flyers and emails about the workshop at informal learning spaces (e.g., public museums, local makerspaces, and after-school program venues).

\textbf{Procedure}. The workshop was divided into four stages, where the first stage consisted of icebreaker activities, the second and third stages focused on constructing and using the toolkit, and the last stage was for concluding and reflection. \autoref{fig:workflow} illustrates the general workflow of the workshop. Before engaging in the activities, the research team first provided an overview of the workshop to the parents and guardians of the children, gained consent from parents, and collected assent from the children. 
We took photographs and collected video-audio recordings as the children were constructing and interacting with the toolkit. 

\textbf{Stage 1. Introduction and Icebreaker Activities.} The workshop facilitators introduced themselves to the children. Children were asked about their perception of data and visualization in general. Afterward, they were shown a series of common visualizations (bar, line, pie) and asked if they recognized the charts. Then children engaged in an open-ended discussion about what they noticed in each of the images to gauge their familiarity with these visualizations more concretely.

\textbf{Stage 2. Toolkit Construction.} In the second stage, children were introduced to \techname. Workshop facilitators first provided a finished toolkit for children to play with and encouraged children to inspect the sliders and mirrors. Children were instructed on how to use the \techname\ by moving the sliders. At this time, children played with the toolkit using the predefined fruit dataset used in the formative study, observing changes displayed on the smartphone. Afterward, children moved on to building their own \techname. Each child was given a cardboard fold-out template and a packet containing the various components required for assembly. Facilitators guided the children through the construction process by demonstrating the physical operations required for assembly. After constructing the kit, children were provided craft materials (e.g., pens, pipe cleaners, stickers) to decorate the outer body of the toolkit.

\textbf{Stage 3. Using the Toolkit}. In the third stage, children used their toolkit to author visualizations. Working under the constraints of a guided and timed workshop, children were asked to generate and collect data using the same game from the formative study for three rounds (c.f., \autoref{sec:formative_methods}). In Round 1, children played the game under five different conditions to author a bar chart. Two of the conditions were predefined by the research team (throwing disks using the left hand and right hand) and the remaining three conditions were defined by the children. In Round 2, from the five conditions, children were asked to repeat one game condition of their choice four more times. This activity was intended to help children generate data to create a line chart. In Round 3, children calculated the total score for all five conditions and visualized each condition’s contribution (i.e., proportion) to the total score for the pie chart.
After finishing the game of disk darts, children were shown how to use the paper templates to translate their collected data into chart axes and other chart elements (e.g., title, legend). They then scanned the templates and authored the different visualizations (i.e., bar, line, pie) using the toolkit. Afterward, children analyzed the resulting graphs and compared their results to others.

\textbf{Stage 4. Conclusion and Reflection.} In the fourth stage, we quickly recapped the activities the children engaged with. Children were asked semi-structured questions to gain their overall impression of the toolkit and data visualization in general. 

\vspace{-2pt}
\subsection{One-on-One Interview with Educators}
\label{sec:teachers_eval}
\textbf{Participants.} We recruited five educators (3 female, 2 male) from a local school district and via our network of education researchers. All educators had experience with teaching visualization to young children in the context of either mathematics or science curricula.
Our aim with the educators was to gain feedback on how \techname\ might be integrated into informal learning spaces. Following the guidelines by Kerzner et al.~\cite{kerzner2019workshops}, we recruited educators with diverse experiences. 

\textbf{Procedure.} Each participant provided a verbal introduction of their background at the start of the interview. The facilitators then performed a demo of the toolkit, illustrating how to build the toolkit and showcasing the different features (e.g., scanning with paper templates, switching panels to create different visualizations). After the demo, educators partook in a concurrent think-aloud protocol where they were encouraged to build and interact with the toolkit in an open-ended session for approximately 20 minutes. Educators then engaged in a semi-structured interview where they answered additional survey questions about their impressions, ideas, and concerns about the toolkit. Educators were audio-recorded and photographed working with our toolkit.

%% file: 6_results.tex
\vspace{-2pt}
\section{Results}
We qualitatively analyzed materials (i.e., notes, images, videos) from our exploratory study. 
We conducted a content analysis on these materials using the design goals that emerged from the formative study as \textit{a priori} categories~\cite{weber1990basic}. Two researchers performed this categorization independently, after which a final set of observations corresponding to each design goal was agreed upon (\autoref{sec:design-goals}). Findings were extracted from these grouped observations. For the interviews with the educators, we summarized their impressions, feedback, and concerns also through the lens of our design goals and research objectives.

\vspace{-2pt} 
\subsection{Workshops with Children}
Children provided positive impressions of the workshop and enjoyed engaging with \techname, despite having greatly varying baseline knowledge of data visualizations (as assessed by the ice-breaker discussions).
In the beginning, the younger children (P1-P4) were not able to assign names to the visualizations, explaining they did not frequently interact with or create visualizations. Yet, they found the workshop particularly fun, with P4 expressing, ``\textit{I really loved it!}'' P3 strongly expressed wanting a version of the toolkit that would work with his iPad, so that he could use it anytime without relying on his parents' smartphones. In contrast, P5, the oldest child of the two workshops, was able to identify, interpret, and explain the function of all charts and found the workshop with the toolkit interesting, but only a ``little fun.'' 

\vspace{-3pt}
\subsubsection{Approachability}
\textbf{G1: Foster Experimentation.} 
Initially, P1-P4 were hesitant to perform certain physical operations, such as folding the flaps of the cardboard base, and even caused minor tears in the cardboard. We then explained to the children that these issues are normal and were, in fact, the reason why we used cardboard and everyday materials: mistakes were a welcome part of the process. After our explanation, P1-P4 began to more confidently interact with the cardboard base and intentionally experiment with different assembly configurations of the components, such as having the phone camera cutout be on the right instead of the left (\autoref{fig:workshop-pics}a). Similarly, with the paper templates, three of the four younger children (P1, P2, P4) initially ripped the template in half when trying to remove the tear-off rectangles. After providing more copies of the template, they were able to successfully author a template. These examples speak to the children's willingness to try again despite the challenges they faced and to experiment with the toolkit in unexpected ways to learn more about the content and function it provided.

\textbf{G2: Replicable.} P1-P4 mentioned they envisioned themselves attempting to recreate the toolkit at home with support from their family. Though children were given pre-packaged kits to construct their own toolkit during the workshop, they provided suggestions on how to further improve the replicability of \techname. P2 and P3 suggested that the overall toolkit should be bigger so it can support larger screen interfaces like a tablet. P3's motivation for this request stemmed from wanting increased agency by using a device he had access to at home. P5, in contrast, felt more comfortable with the idea of constructing a toolkit by herself and specifically envisioned how the toolkit could be used by children of her age at her school's library. She explained how the school librarian would often make children do activities related to data visualization (e.g., create a pie chart to track genres of books read), and that the toolkit might make for an interesting way to author multiple charts related to that information. \techname's design showcases the breadth of how different children envision the toolkit being reused in different contexts.

\vspace{-2pt}
\subsubsection{Engagement}
\textbf{G3: Ease of Use.}
In both workshops, we provided verbal instructions on how to construct the box and also demonstrated relevant actions in real-time. 
The components of the toolkit were designed so that it would be easy for children to assemble the box by themselves. 
This was true for P5, but the younger participants all required more time and some hands-on assistance to complete the process. The younger children found it difficult to securely insert some cardboard flaps into the pre-cut slots especially when the pieces were not already connected. They also needed instructions while arranging fiducial markers on the back of the panels in the right order. P5 took 15 minutes to complete the box assembly, while P1-P4 needed around 35 minutes.

\begin{figure}[tb]
	\centering
	\captionsetup{farskip=0pt}
	\includegraphics[width=\linewidth]{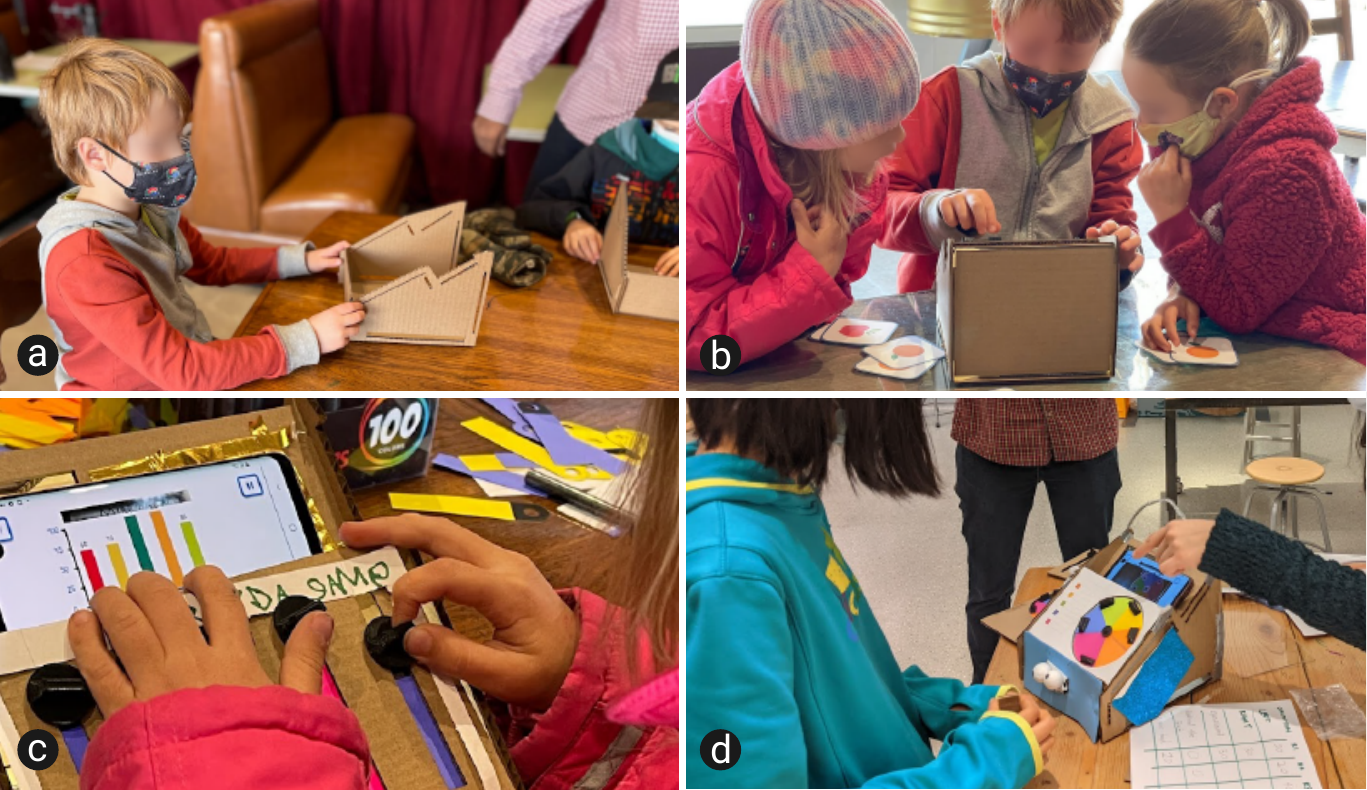}
    \caption{(a) P3 assembling the cardboard base (b) P1--P3 collaborating (c) P1 interacting with her authored bar chart (d) P5's decorated toolkit.}
    \vspace{-0.2cm}
	\label{fig:workshop-pics}
\end{figure}

Despite this variance, \techname\ successfully enabled all children to transform their data into digital charts. All participants were able to successfully create visualizations even from the workshop's first activity: translate sorted stacks of fruit images into a digital bar chart using the toolkit slider. The interface also encouraged the participants to construct the kit by themselves and promoted collaboration. In the group workshop, despite no instructions to work together, P1-P3 started to naturally collaborate to help each other create and provide comments on each other's bar chart (\autoref{fig:workshop-pics}b).

\textbf{G4: Support creative expression.}
After constructing the toolkit, children personalized their box using a range of craft materials including sketch pens, colored paper, and stickers. Children particularly enjoyed this step, spending significant time (20 minutes on average) deciding which materials to use and how to design their box. The younger children began by writing their names on the box, adding stickers, coloring the sides, and drawing faces on the box to personify it. P5, who had the greatest familiarity with visualizations, focused on many functional additions, such as textual instructions for where to place the mobile phone, a handle to carry the box made using pipe cleaners that were inserted into the flutes of the cardboard, and two small knobs to keep the paper template in place (\autoref{fig:workshop-pics}d).

Children's creative expression was not limited to solely the box but also included the paper templates and paper strips for the bar charts. With the paper templates, children came up with a range of titles for their visualizations, such as ``Frisbee,'' ``Throw-it Game,'' ``[P3's] Game,'' and ``Targeting.'' Children also exhibited creative variations in how they labeled their x-axes: three children expressed the x-axis labels with full words (\autoref{fig:workshop-pics}c), one used a combination of letters and numbers, and one drew icons. With the paper strips, children used three types of color schemes: all of the same colors (P3--to use their favorite color), all different colors (P1, P2, P5--to make use of all available options), or two alternating colors (P4--to imitate the facilitator's toolkit). Even the simple decisions of choosing paper strips, deciding titles, and naming x-axis labels highlight how children were excited to showcase their creative expressions.

\vspace{-3pt}
\subsubsection{Scaffolding \& Embodied Learning}
\textbf{G5: Craft visualizations.} 
Although children faced some issues during the construction process, most children were able to create, analyze and draw meaning from their visualizations. Across both workshops, children spent about 30 minutes building and interacting with the chart panels. Towards the end of the workshop, P2 and P5 inferred from their charts that their performance with one hand was better than the other due to better control of their dominant hand. 
P5 noticed a slight upward trend in her line chart as she performed better over time as she played the game. She elaborated that the resemblance of the physical line chart to its digital form drew her attention to the line's shape. Similarly, P1-P4 were also able to draw parallels between the physical bar chart and the digital representation. While playing with the bar chart control panels (moving sliders up and down), they noted how the height of the paper strips proportionally matched the heights of the digital bars. The ability to physically craft the visualization and the manner in which the digital representation mirrored the movement of the 3D printed sliders helped reinforce the various visualization components that they interacted with.

\textbf{G6: Physical Interactions.} After interacting with the facilitators' toolkit, children immediately raised questions about and tried to explain the box's mechanism. P3 hypothesized the mirrors were necessary for the camera to see fiducial markers ``...to provide some kind of coding that's connect[ing] the markers and phone.'' P3 continued to question the toolkit's mechanism, asking ``\textit{So basically you build the data [visualization] with [the toolkit], and then you print [the visualization] out?}'' This curiosity and fascination motivated the younger children to take the lead in the learning and creation process. For example, P1-P4 expressed great excitement in building their own \techname\ and chose to construct the toolkit before playing the game of disk darts (\autoref{fig:workflow}b) when offered a choice. Though the construction process took longer for the younger children, they were engaged throughout the entire process. They continuously questioned how certain components and features were related to data visualizations. For example, P4 asked, ``\textit{Why are there these slots when we aren't using them?}'' when he noticed the two slots at the base of the box that were meant to let in light to the bottom mirror. Similarly, P2 questioned why there were multiple colors for the paper strips and why it mattered for visualization.

While the toolkit fostered their curiosity, children also noted improvements in the toolkit's construction process. P5 wished that the toolkit box would be more stable while she was decorating her box. Additionally, while constructing visualizations, she noticed that the sliders did not smoothly move for all panels (bar, line, pie). Consequently, for the bar and line charts, she spent approximately three minutes rapidly moving the sliders up and down the channels in an attempt to manually smooth the cardboard structure. P1-P4 faced a bottleneck during the panel assembly process when they tried to affix the two parts of the sliders to the chart panel. The grooves within some of the sliders were worn down after a few cycles of opening and closing, and the younger children requested assistance during this step.

\begin{figure*}[t]
	\centering
	\captionsetup{farskip=0pt}
	\includegraphics[width=\linewidth]{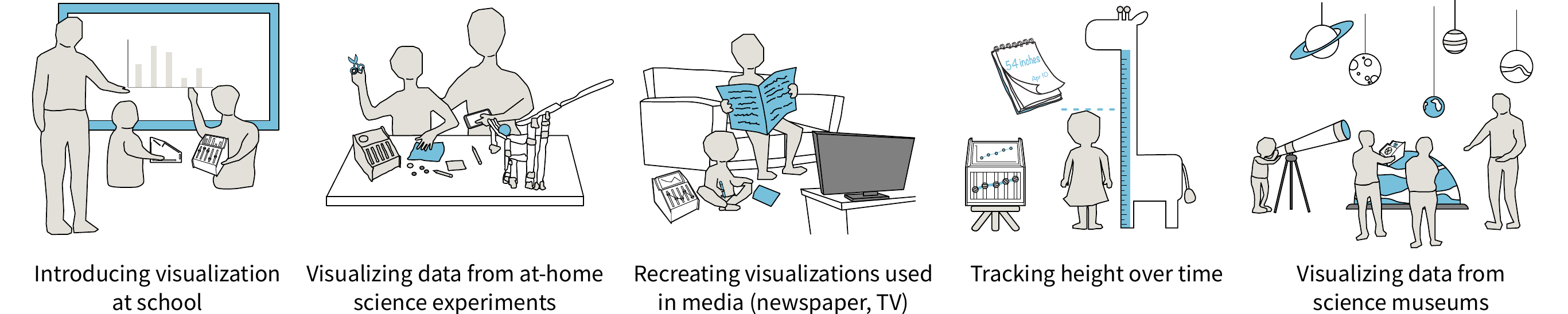}
    \caption{Sketches of incorporating \techname\ into various scenarios to expand ways children engage with data.}
    \vspace{-0.1cm}
	\label{fig:future-vision}
\end{figure*}

\vspace{-2pt}
\subsection{Interview with Educators}
\vspace{-2pt}
Educators were overall appreciative of \techname. E1, E2, and E5 specifically appreciated the DIY aspect of the toolkit. E1's and E2's after-school programs emphasize making and physical activities, and they noted the toolkit was a clever solution to introduce children to data visualizations in line with those pedagogical goals. All three educators also felt that the ability to personalize both the overall appearance of the box and the charts can help children better identify with the activities and take more ownership of the results. As a researcher for early childhood education, E4 emphasized the importance of physical manipulation for young children's education, pointing out that even adults use physical operations to help them think more clearly (e.g., using their hands to count). Based on these comments, E4 noted \techname\ could help students have another mental metaphor to work with when reasoning about abstract concepts in data.

However, both E3 and E5 commented on the toolkit's limitations concerning the mobile phone. While E5 commented that the toolkit is a powerful introduction for younger children (i.e., 6-8 years old), ``\textit{the [current] toolkit is going to [reach] a ceiling where [children] have already reached to highest capability they can go [with the phone].}'' Similarly, E3  questioned the choice of a mobile phone. She felt smartphones provide a useful introduction to the common visualizations but will be quickly limited in their ability to create complex charts later on. She suggested a similar toolkit scheme could incorporate a tablet or laptop in the future. 
We assume that these concerns arise from the perceived limitation of the compact physical form in addition to the mobile phone's capability relative to laptops when visualizing data. Both educators found it difficult to imagine how the toolkit can incorporate more than five data points in its current form while retaining sufficiently large marks for children to work with.

For future work, E3--E5 stressed how \techname\ could be integrated into existing curriculum (e.g., connecting with schools or after-school programs). They commented the strength of \techname\ stems from how easily it can be integrated into existing activities. E1 and E4 suggested integrating \techname\ into informal learning spaces that are likely to already have their own instructional content relating to visualizations (e.g., science museums).
E1 explained that in her past experience as a middle school teacher, she frequently asked her students to record their daily homework progress using stickers in a notebook, similar to the process of creating and saving charts in Flow 3. E1 also expressed the challenges she experienced when students misunderstood or were unwilling to create visualizations for their science lab reports. She stated a toolkit like \techname\ might have helped her students with those two activities more easily, while also resulting in engaging, informative visualizations throughout the school year.

%% file: 7_discussion.tex
\vspace{-2pt}
\section{Discussion}
We used \techname\ as a design probe to understand two core research questions: (1) how might children respond to physical crafting experiences in the context of working with visualizations and (2) how can we expand DVL learning to informal learning spaces?
We found that \techname\ helped younger children learn more about visualization components, while older children who were more familiar with data visualizations found \techname\ to be more of an engaging authoring tool (\autoref{sec:data-is-yours}). 
Educators immediately saw the benefits of our approach and how it could be expanded and integrated with other activities in informal learning spaces
(\autoref{sec:research-agenda}).

\vspace{-2pt}
\subsection{Data is Yours}
\label{sec:data-is-yours}
Integrating physical crafting into a visualization authoring context was useful for children but that utility differed based on the children's age and prior knowledge of data visualization. 

Older children, who are likely to be more familiar with visualizations, viewed the toolkit primarily as a platform for creating their own visualizations with more agency. At the end of the workshop, P5, the oldest child, clearly stated she did not learn anything new about data visualization from the workshop. However, she noted how \techname\ was a more engaging method of authoring charts that she could see herself using in her daily life. Although she had used spreadsheet programs like Google Sheets to visualize her reading habits as pie charts,  she described the limitations of the ``black box'' nature of purely digital tools: ``\textit{[With Google sheets] you just type in a bunch of numbers and [the visualization] appears, but with \techname\ I can actually create charts with my friends and talk about it.}'' P5 also spoke positively of how she could use \techname\ and the paper templates to author charts using any dataset (e.g., exercise time, how long she spends coding). P5's statements support the idea that the physical and interactive nature of the \techname\ results in a more engaging and empowering process compared to exclusively digital visualization software. 

\vspace{-2pt}
\subsection{Limitations \& Future Work}
Our exploratory study does not aim to provide any direct pedagogical evidence of our approach improving children's DVL. Rather the qualitative observations we collected provide preliminary evidence that the children enjoyed interacting with data physically, can familiarize themselves with foundational concepts of visualization (e.g., reason about visual semantics, incorporate chart elements, set up appropriate scales) through curiosity and play, and were empowered to creatively engage with data. 
We currently lack more formal assessment methods that are appropriate for children, such as revised visualization literacy assessment tests, that would allow for a more formal DVL measurement. 

Though \techname\ is designed to help children author visualizations using data of their choice, the tool's expressiveness was not fully explored within the constraints of a two-hour workshop. 
For example, the importance of working with personally meaningful datasets varied based on age. P1-P4 did not mind interacting with the predefined fruit datasets. P5 expressed a similar sentiment as the two formative children, who were similarly aged, indicating that the fruit datasets felt childish and the self-generated game datasets were more enjoyable. 

Additionally, a majority of the time in the workshop was dedicated to assembling \techname. While further observations of how children engaged with their visualizations would have provided more in-depth insights, we found that assembling \techname\ helped children accumulate playful learning experiences that sparked an interest in and new comfort with visualization.
Extensive DVL encompasses complex knowledge and reasoning processes that go beyond the two-hour activities described here.
Thus, further studies will be required to see how children will use \techname\ to author data visualizations in an unconstrained and longitudinal setting. 

Assessing children's DVL frequently focuses on reading, creating, and interpreting visualizations (\autoref{sec:toolkit}). However, DVL also encompasses critical engagement with visualizations, such as understanding when charts are misleading~\cite{camba2022deception, laurer2020}. Our toolkit currently requires facilitator feedback to let children know about misleading aspects of their visualizations. While we did not observe such errors, they are possible given the open-ended authoring functionality (e.g., using misleading axis values). To develop critical design literacy, future solutions could incorporate visual linting and similar verification strategies~\cite{hopkins2020visualint}.

\vspace{-2pt}
\subsection{A Research Agenda for Children's DVL}
\label{sec:research-agenda}
With any learning technology for children, research stresses the importance of three principles: ``low floor'' (how easy is the technology for children to learn and use?); ``high ceiling'' (does the technology advance children in making more complex projects?); and ``wide walls'' (how inclusive is the technology in supporting all children's interests?)~\cite{resnick2005design}.
Our observations from the workshops indicate that \techname\ successfully accomplished a \textit{low floor}. Young, inexperienced children were able to craft and interpret basic visualizations using the system. But our approach requires more consideration as to how to create a \textit{high ceiling}. E3 and E5 expressed concerns about the toolkit's current dependency on a smartphone. However, as children become older (e.g., teenagers), we anticipate they will not need the scaffolding from the physical panels to learn and interact with visualizations. Rather, research shows various ways to interact with data that are particularly suited for mobile devices~\cite{brehmer2019, brehmer2021interacting}. E3's and E5's comments indicate that future iterations of the toolkit could benefit from further considerations on how to better leverage state-of-the-art visualization techniques to engage older children.

Expanding \techname\ to additional contexts would also increase the variety and complexity of charts children can create. 
Children and educators suggested several ways the toolkit can support \textit{wide walls}, including integrating the toolkit into activities that children naturally enjoy (\autoref{fig:future-vision}).
For instance, children can have a more sustained engagement with data at home by working with their family to recreate visualizations from news media or progressively tracking changes around them over time (e.g., tracking children's height). These activities, in turn, can seamlessly introduce children to different types of data (e.g., qualitative, geographic, categorical, temporal). 
\techname\ can also be incorporated into group activities that already take place in informal learning spaces (e.g., science museums). The constructionist approach shows how produced artifacts can be ``shown, discussed, probed, and admired''~\cite{papert1993children}, serving as a locus for collaborative discussions. These actions align with our observations from the group workshop where children collaborated with each other and provided feedback on their peers' visualizations without being prompted to. 
We believe expanding \techname\ into daily contexts and collaborative settings can help children be more critical of the data found around them.

%% file: 8_conclusion.tex
\vspace{-2pt}
\section{Conclusion}
We explored how a constructionist approach can foster children's data visualization literacy (DVL) using \techname.
Our observations from a series of workshops with children ($n=5$; 6--11 years old) and interviews with educators ($n=5$) reveal that physical crafting with visualizations benefits children in different ways depending on their past experiences with data. This exploratory study confirms the need to further investigate how to expand the ways children can interact with data and visualizations. 
We reflect on findings and current research gaps to identify new challenges in children's DVL research. 
We hope that our approach can inspire new ways for children to creatively and playfully interact with the visualizations and data that are woven into the fabrics of everyday life.

%% file: 9_acknowledgement.tex
\acknowledgments{
The authors would like to thank Dr. Shaz Zamore for their assistance on the project.
This material is based upon work supported by the National Science Foundation under Grant No. IIS-2040489, IIS-2046725, \& STEM+C 1933915. }

%% file: 00_main.bbl
\begin{thebibliography}{10}

\bibitem{airinei2010data}
D.~Airinei and D.~Homocianu.
\newblock Data visualization in business intelligence.
\newblock {\em 2010 Proceeding of WSEAS MCBEC 2010 - Recent Advances in
  Mathematics and Computers in Business, Economics, Biology \& Chemistry},
  2010.

\bibitem{alimisis2009constructionism}
D.~Alimisis and C.~Kynigos.
\newblock Constructionism and robotics in education.
\newblock {\em Teacher education on robotic-enhanced constructivist pedagogical
  methods}, pp. 11--26, 2009.

\bibitem{alper2017visualization}
B.~Alper, N.~H. Riche, F.~Chevalier, J.~Boy, and M.~Sezgin.
\newblock Visualization literacy at elementary school.
\newblock In {\em Proceedings of the 2017 CHI Conference on Human Factors in
  Computing Systems}, CHI '17, p. 5485–5497. Association for Computing
  Machinery, New York, NY, USA, 2017. doi: {{%
10\hspace{.1pt}\discretionary{.}{%
}{.}\hspace{.4pt}1145\discretionary{/}{%
}{/}3025453\hspace{.1pt}\discretionary{.}{%
}{.}\hspace{.4pt}3025877}}


\bibitem{anderson2012eroding}
M.~L. Anderson, M.~J. Richardson, and A.~Chemero.
\newblock Eroding the boundaries of cognition: Implications of embodiment 1.
\newblock {\em Topics in cognitive science}, 4(4):717--730, 2012.

\bibitem{bae2021idc}
S.~Bae, R.~Yang, P.~Gyory, J.~Uhr, D.~A. Szafir, and E.~Y.-L. Do.
\newblock Touching information with diy paper charts \& ar markers.
\newblock In {\em Interaction Design and Children}, IDC '21, p. 433–438.
  Association for Computing Machinery, New York, NY, USA, 2021. doi: {{%
10\hspace{.1pt}\discretionary{.}{%
}{.}\hspace{.4pt}1145\discretionary{/}{%
}{/}3459990\hspace{.1pt}\discretionary{.}{%
}{.}\hspace{.4pt}3465191}}


\bibitem{bhargava2017data}
R.~Bhargava and C.~D'Ignazio.
\newblock Data sculptures as a playful and low-tech introduction to working
  with data.
\newblock In {\em Proceedings of the 2017 Designing Interactive Systems}.
  Association for Computing Machinery, 2017.

\bibitem{bishop2020construct}
F.~Bishop, J.~Zagermann, U.~Pfeil, G.~Sanderson, H.~Reiterer, and U.~Hinrichs.
\newblock Construct-a-vis: Exploring the free-form visualization processes of
  children.
\newblock {\em IEEE Transactions on Visualization and Computer Graphics},
  26(1):451--460, 2020. doi: {{%
10\hspace{.1pt}\discretionary{.}{%
}{.}\hspace{.4pt}1109\discretionary{/}{%
}{/}TVCG\hspace{.1pt}\discretionary{.}{%
}{.}\hspace{.4pt}2019\hspace{.1pt}\discretionary{.}{%
}{.}\hspace{.4pt}2934804}}


\bibitem{bohman2015data}
S.~Bohman.
\newblock Data visualization: an untapped potential for political participation
  and civic engagement.
\newblock In {\em International Conference on Electronic Government and the
  Information Systems Perspective}, pp. 302--315. Springer, 2015.

\bibitem{bostockd3}
M.~Bostock, V.~Ogievetsky, and J.~Heer.
\newblock D³ data-driven documents.
\newblock {\em IEEE Transactions on Visualization and Computer Graphics},
  17(12):2301--2309, 2011. doi: {{%
10\hspace{.1pt}\discretionary{.}{%
}{.}\hspace{.4pt}1109\discretionary{/}{%
}{/}TVCG\hspace{.1pt}\discretionary{.}{%
}{.}\hspace{.4pt}2011\hspace{.1pt}\discretionary{.}{%
}{.}\hspace{.4pt}185}}


\bibitem{boy2014principled}
J.~Boy, R.~A. Rensink, E.~Bertini, and J.-D. Fekete.
\newblock A principled way of assessing visualization literacy.
\newblock {\em IEEE Transactions on Visualization and Computer Graphics},
  20(12):1963--1972, 2014. doi: {{%
10\hspace{.1pt}\discretionary{.}{%
}{.}\hspace{.4pt}1109\discretionary{/}{%
}{/}TVCG\hspace{.1pt}\discretionary{.}{%
}{.}\hspace{.4pt}2014\hspace{.1pt}\discretionary{.}{%
}{.}\hspace{.4pt}2346984}}


\bibitem{brehmer2019}
M.~Brehmer, B.~Lee, P.~Isenberg, and E.~K. Choe.
\newblock A comparative evaluation of animation and small multiples for trend
  visualization on mobile phones.
\newblock {\em IEEE Transactions on Visualization and Computer Graphics},
  26(1):364--374, 2020. doi: {{%
10\hspace{.1pt}\discretionary{.}{%
}{.}\hspace{.4pt}1109\discretionary{/}{%
}{/}TVCG\hspace{.1pt}\discretionary{.}{%
}{.}\hspace{.4pt}2019\hspace{.1pt}\discretionary{.}{%
}{.}\hspace{.4pt}2934397}}


\bibitem{brehmer2021interacting}
M.~Brehmer, B.~Lee, J.~Stasko, and C.~Tominski.
\newblock Interacting with visualization on mobile devices.
\newblock In {\em Mobile Data Visualization}, pp. 67--110. Chapman and
  Hall/CRC, 2021.

\bibitem{borner2019framework}
K.~Börner, A.~Bueckle, and M.~Ginda.
\newblock Data visualization literacy: Definitions, conceptual frameworks,
  exercises, and assessments.
\newblock {\em Proceedings of the National Academy of Sciences},
  116(6):1857--1864, 2019. doi: {{%
10\hspace{.1pt}\discretionary{.}{%
}{.}\hspace{.4pt}1073\discretionary{/}{%
}{/}pnas\hspace{.1pt}\discretionary{.}{%
}{.}\hspace{.4pt}1807180116}}


\bibitem{camba2022deception}
J.~D. Camba, P.~Company, and V.~L. Byrd.
\newblock Identifying deception as a critical component of visualization
  literacy.
\newblock {\em IEEE Computer Graphics and Applications}, 42(1):116--122, 2022.
  doi: {{%
10\hspace{.1pt}\discretionary{.}{%
}{.}\hspace{.4pt}1109\discretionary{/}{%
}{/}MCG\hspace{.1pt}\discretionary{.}{%
}{.}\hspace{.4pt}2021\hspace{.1pt}\discretionary{.}{%
}{.}\hspace{.4pt}3132004}}


\bibitem{Chevalier2018visualization}
F.~Chevalier, N.~Henry~Riche, B.~Alper, C.~Plaisant, J.~Boy, and N.~Elmqvist.
\newblock Observations and reflections on visualization literacy in elementary
  school.
\newblock {\em IEEE Computer Graphics and Applications}, 38(3):21--29, 2018.
  doi: {{%
10\hspace{.1pt}\discretionary{.}{%
}{.}\hspace{.4pt}1109\discretionary{/}{%
}{/}MCG\hspace{.1pt}\discretionary{.}{%
}{.}\hspace{.4pt}2018\hspace{.1pt}\discretionary{.}{%
}{.}\hspace{.4pt}032421650}}


\bibitem{claes2015}
S.~Claes and A.~V. Moere.
\newblock The role of tangible interaction in exploring information on public
  visualization displays.
\newblock In {\em Proceedings of the 4th International Symposium on Pervasive
  Displays}, PerDis '15, p. 201–207. Association for Computing Machinery, New
  York, NY, USA, 2015. doi: {{%
10\hspace{.1pt}\discretionary{.}{%
}{.}\hspace{.4pt}1145\discretionary{/}{%
}{/}2757710\hspace{.1pt}\discretionary{.}{%
}{.}\hspace{.4pt}2757733}}


\bibitem{dasgupta2017scratch}
S.~Dasgupta and B.~M. Hill.
\newblock Scratch community blocks: Supporting children as data scientists.
\newblock In {\em Proceedings of the 2017 CHI Conference on Human Factors in
  Computing Systems}, CHI '17, p. 3620–3631. Association for Computing
  Machinery, New York, NY, USA, 2017. doi: {{%
10\hspace{.1pt}\discretionary{.}{%
}{.}\hspace{.4pt}1145\discretionary{/}{%
}{/}3025453\hspace{.1pt}\discretionary{.}{%
}{.}\hspace{.4pt}3025847}}


\bibitem{d2016databasic}
C.~D'Ignazio and R.~Bhargava.
\newblock Databasic: Design principles, tools and activities for data literacy
  learners.
\newblock {\em The Journal of Community Informatics}, 12(3), 2016.

\bibitem{d2018creative}
C.~D'Ignazio and R.~Bhargava.
\newblock Creative data literacy: A constructionist approach to teaching
  information visualization.
\newblock {\em Digital Humanities Quarterly}, 2018.

\bibitem{diseases2020covid}
T.~L.~I. Diseases.
\newblock The covid-19 infodemic.
\newblock {\em The Lancet. Infectious Diseases}, 20(8):875, 2020.

\bibitem{doan2021misrepresenting}
S.~Doan.
\newblock Misrepresenting covid-19: lying with charts during the second golden
  age of data design.
\newblock {\em Journal of Business and Technical Communication}, 35(1):73--79,
  2021.

\bibitem{dork2013critical}
M.~D{\"o}rk, P.~Feng, C.~Collins, and S.~Carpendale.
\newblock Critical infovis: exploring the politics of visualization.
\newblock In {\em CHI'13 Extended Abstracts on Human Factors in Computing
  Systems}, pp. 2189--2198. Association for Computing Machinery, New York, NY,
  USA, 2013.

\bibitem{druin2002role}
A.~Druin.
\newblock The role of children in the design of new technology.
\newblock {\em Behaviour and information technology}, 21(1):1--25, 2002.

\bibitem{fekete2004toolkit}
J.-D. Fekete.
\newblock The infovis toolkit.
\newblock In {\em IEEE Symposium on Information Visualization}, pp. 167--174,
  2004. doi: {{%
10\hspace{.1pt}\discretionary{.}{%
}{.}\hspace{.4pt}1109\discretionary{/}{%
}{/}INFVIS\hspace{.1pt}\discretionary{.}{%
}{.}\hspace{.4pt}2004\hspace{.1pt}\discretionary{.}{%
}{.}\hspace{.4pt}64}}


\bibitem{nces2021}
N.~C. for Educational~Statistics.
\newblock Children’s internet access at home, 2021.
\newblock \url{https://nces.ed.gov/programs/coe/indicator/cch}.

\bibitem{gabler2019safari}
J.~G\"{a}bler, C.~Winkler, N.~Lengyel, W.~Aigner, C.~Stoiber, G.~Wallner, and
  S.~Kriglstein.
\newblock Diagram safari: A visualization literacy game for young children.
\newblock In {\em Extended Abstracts of the Annual Symposium on Computer-Human
  Interaction in Play Companion Extended Abstracts}, CHI PLAY '19 Extended
  Abstracts, p. 389–396. Association for Computing Machinery, New York, NY,
  USA, 2019. doi: {{%
10\hspace{.1pt}\discretionary{.}{%
}{.}\hspace{.4pt}1145\discretionary{/}{%
}{/}3341215\hspace{.1pt}\discretionary{.}{%
}{.}\hspace{.4pt}3356283}}


\bibitem{garrido2016generation}
S.~Garrido-Jurado, R.~Muñoz-Salinas, F.~Madrid-Cuevas, and R.~Medina-Carnicer.
\newblock Generation of fiducial marker dictionaries using mixed integer linear
  programming.
\newblock {\em Pattern Recognition}, 51:481--491, 2016. doi: {{%
10\hspace{.1pt}\discretionary{.}{%
}{.}\hspace{.4pt}1016\discretionary{/}{%
}{/}j\hspace{.1pt}\discretionary{.}{%
}{.}\hspace{.4pt}patcog\hspace{.1pt}\discretionary{.}{%
}{.}\hspace{.4pt}2015\hspace{.1pt}\discretionary{.}{%
}{.}\hspace{.4pt}09\hspace{.1pt}\discretionary{.}{%
}{.}\hspace{.4pt}023}}


\bibitem{goswami2002blackwell}
U.~E. Goswami.
\newblock {\em Blackwell handbook of childhood cognitive development.}
\newblock Blackwell publishing, 2002.

\bibitem{gourlet2017cairn}
P.~Gourlet and T.~Dass\'{e}.
\newblock Cairn: A tangible apparatus for situated data collection,
  visualization and analysis.
\newblock In {\em Proceedings of the 2017 Conference on Designing Interactive
  Systems}, DIS '17, p. 247–258. Association for Computing Machinery, New
  York, NY, USA, 2017. doi: {{%
10\hspace{.1pt}\discretionary{.}{%
}{.}\hspace{.4pt}1145\discretionary{/}{%
}{/}3064663\hspace{.1pt}\discretionary{.}{%
}{.}\hspace{.4pt}3064794}}


\bibitem{heer2005prefuse}
J.~Heer, S.~K. Card, and J.~A. Landay.
\newblock Prefuse: A toolkit for interactive information visualization.
\newblock In {\em Proceedings of the SIGCHI Conference on Human Factors in
  Computing Systems}, CHI '05, p. 421–430. Association for Computing
  Machinery, New York, NY, USA, 2005. doi: {{%
10\hspace{.1pt}\discretionary{.}{%
}{.}\hspace{.4pt}1145\discretionary{/}{%
}{/}1054972\hspace{.1pt}\discretionary{.}{%
}{.}\hspace{.4pt}1055031}}


\bibitem{pasek2015learners}
K.~Hirsh-Pasek, J.~M. Zosh, R.~M. Golinkoff, J.~H. Gray, M.~B. Robb, and
  J.~Kaufman.
\newblock Putting education in “educational” apps: Lessons from the science
  of learning.
\newblock {\em Psychological Science in the Public Interest}, 16(1):3--34,
  2015. doi: {{%
10\hspace{.1pt}\discretionary{.}{%
}{.}\hspace{.4pt}1177\discretionary{/}{%
}{/}1529100615569721}}


\bibitem{hopkins2020craft}
A.~Hopkins, N.~A. Vladis, and A.~Satyanarayan.
\newblock Data crafting: Exploring data through craft and play.
\newblock In {\em VisActivities: IEEE VIS Workshop on Data Vis Activities at
  IEEE VIS 2020}, 2020.

\bibitem{hopkins2020visualint}
A.~K. Hopkins, M.~Correll, and A.~Satyanarayan.
\newblock Visualint: Sketchy in situ annotations of chart construction errors.
\newblock In {\em Computer Graphics Forum}, vol.~39, pp. 219--228. Wiley Online
  Library, 2020.

\bibitem{huron2016using}
S.~Huron, S.~Carpendale, J.~Boy, and J.-D. Fekete.
\newblock Using viskit: A manual for running a constructive visualization
  workshop.
\newblock In {\em Pedagogy of Data Visualization Workshop at IEEE VIS 2016},
  2016.

\bibitem{huron2014constructiveviz}
S.~Huron, S.~Carpendale, A.~Thudt, A.~Tang, and M.~Mauerer.
\newblock Constructive visualization.
\newblock In {\em Proceedings of the 2014 Conference on Designing Interactive
  Systems}, DIS '14, p. 433–442. Association for Computing Machinery, New
  York, NY, USA, 2014. doi: {{%
10\hspace{.1pt}\discretionary{.}{%
}{.}\hspace{.4pt}1145\discretionary{/}{%
}{/}2598510\hspace{.1pt}\discretionary{.}{%
}{.}\hspace{.4pt}2598566}}


\bibitem{huron2017workshop}
S.~Huron, P.~Gourlet, U.~Hinrichs, T.~Hogan, and Y.~Jansen.
\newblock Let's get physical: Promoting data physicalization in workshop
  formats.
\newblock In {\em Proceedings of the 2017 Conference on Designing Interactive
  Systems}, DIS '17, p. 1409–1422. Association for Computing Machinery, New
  York, NY, USA, 2017. doi: {{%
10\hspace{.1pt}\discretionary{.}{%
}{.}\hspace{.4pt}1145\discretionary{/}{%
}{/}3064663\hspace{.1pt}\discretionary{.}{%
}{.}\hspace{.4pt}3064798}}


\bibitem{huron2014constructingtokens}
S.~Huron, Y.~Jansen, and S.~Carpendale.
\newblock Constructing visual representations: Investigating the use of
  tangible tokens.
\newblock {\em IEEE Transactions on Visualization and Computer Graphics},
  20(12):2102--2111, 2014. doi: {{%
10\hspace{.1pt}\discretionary{.}{%
}{.}\hspace{.4pt}1109\discretionary{/}{%
}{/}TVCG\hspace{.1pt}\discretionary{.}{%
}{.}\hspace{.4pt}2014\hspace{.1pt}\discretionary{.}{%
}{.}\hspace{.4pt}2346292}}


\bibitem{huynh2020rpg}
E.~Huynh, A.~Nyhout, P.~Ganea, and F.~Chevalier.
\newblock Designing narrative-focused role-playing games for visualization
  literacy in young children.
\newblock {\em IEEE Transactions on Visualization and Computer Graphics},
  27(2):924--934, 2021. doi: {{%
10\hspace{.1pt}\discretionary{.}{%
}{.}\hspace{.4pt}1109\discretionary{/}{%
}{/}TVCG\hspace{.1pt}\discretionary{.}{%
}{.}\hspace{.4pt}2020\hspace{.1pt}\discretionary{.}{%
}{.}\hspace{.4pt}3030464}}


\bibitem{js-port}
jcmellado.
\newblock Javascript port of the aruco libray.
\newblock \url{https://github.com/jcmellado/js-aruco}, 2018.

\bibitem{jones2017data}
A.~M. Jones.
\newblock Data visualization and health econometrics.
\newblock {\em Foundations and Trends in Econometrics}, 2017.

\bibitem{kahn2021constructionism}
K.~Kahn and N.~Winters.
\newblock Constructionism and ai: A history and possible futures.
\newblock {\em British Journal of Educational Technology}, 52(3):1130--1142,
  2021.

\bibitem{kerzner2019workshops}
E.~Kerzner, S.~Goodwin, J.~Dykes, S.~Jones, and M.~Meyer.
\newblock A framework for creative visualization-opportunities workshops.
\newblock {\em IEEE Transactions on Visualization and Computer Graphics},
  25(1):748--758, 2019. doi: {{%
10\hspace{.1pt}\discretionary{.}{%
}{.}\hspace{.4pt}1109\discretionary{/}{%
}{/}TVCG\hspace{.1pt}\discretionary{.}{%
}{.}\hspace{.4pt}2018\hspace{.1pt}\discretionary{.}{%
}{.}\hspace{.4pt}2865241}}


\bibitem{knudesen2018}
S.~Knudsen, J.~Vermeulen, D.~Kosminsky, J.~Walny, M.~West, C.~Frisson,
  B.~Adriel~Aseniero, L.~MacDonald~Vermeulen, C.~Perin, L.~Quach, P.~Buk,
  K.~Tabuli, S.~Chopra, W.~Willett, and S.~Carpendale.
\newblock Democratizing open energy data for public discourse using
  visualization.
\newblock In {\em Extended Abstracts of the 2018 CHI Conference on Human
  Factors in Computing Systems}, CHI EA '18, p. 1–4. Association for
  Computing Machinery, New York, NY, USA, 2018. doi: {{%
10\hspace{.1pt}\discretionary{.}{%
}{.}\hspace{.4pt}1145\discretionary{/}{%
}{/}3170427\hspace{.1pt}\discretionary{.}{%
}{.}\hspace{.4pt}3186539}}


\bibitem{kokina2017role}
J.~Kokina, D.~Pachamanova, and A.~Corbett.
\newblock The role of data visualization and analytics in performance
  management: Guiding entrepreneurial growth decisions.
\newblock {\em Journal of Accounting Education}, 38:50--62, 2017.

\bibitem{laurer2020}
C.~Lauer and S.~O'Brien.
\newblock How people are influenced by deceptive tactics in everyday charts and
  graphs.
\newblock {\em IEEE Transactions on Professional Communication},
  63(4):327--340, 2020. doi: {{%
10\hspace{.1pt}\discretionary{.}{%
}{.}\hspace{.4pt}1109\discretionary{/}{%
}{/}TPC\hspace{.1pt}\discretionary{.}{%
}{.}\hspace{.4pt}2020\hspace{.1pt}\discretionary{.}{%
}{.}\hspace{.4pt}3032053}}


\bibitem{2021-viral-visualizations}
C.~Lee, T.~Yang, G.~Inchoco, G.~M. Jones, and A.~Satyanarayan.
\newblock {Viral Visualizations: How Coronavirus Skeptics Use Orthodox Data
  Practices to Promote Unorthodox Science Online}.
\newblock In {\em ACM Human Factors in Computing Systems (CHI)}, 2021. doi: {{%
10\hspace{.1pt}\discretionary{.}{%
}{.}\hspace{.4pt}1145\discretionary{/}{%
}{/}3411764\hspace{.1pt}\discretionary{.}{%
}{.}\hspace{.4pt}3445211}}


\bibitem{lee2017vlat}
S.~Lee, S.-H. Kim, and B.~C. Kwon.
\newblock Vlat: Development of a visualization literacy assessment test.
\newblock {\em IEEE Transactions on Visualization and Computer Graphics},
  23(1):551--560, 2017. doi: {{%
10\hspace{.1pt}\discretionary{.}{%
}{.}\hspace{.4pt}1109\discretionary{/}{%
}{/}TVCG\hspace{.1pt}\discretionary{.}{%
}{.}\hspace{.4pt}2016\hspace{.1pt}\discretionary{.}{%
}{.}\hspace{.4pt}2598920}}


\bibitem{lee2022data}
V.~R. Lee and V.~Delaney.
\newblock Identifying the content, lesson structure, and data use within
  pre-collegiate data science curricula.
\newblock {\em Journal of Science Education and Technology}, 31(1):81--98, Feb
  2022. doi: {{%
10\hspace{.1pt}\discretionary{.}{%
}{.}\hspace{.4pt}1007\discretionary{/}{%
}{/}s10956\discretionary{%
}{-}{-}021\discretionary{%
}{-}{-}09932\discretionary{%
}{-}{-}1}}


\bibitem{electronics10050604}
L.~Liu, J.~Guo, C.~Zhang, Z.~Wang, P.~Zhu, T.~Fang, J.~Wang, C.~Yao, and
  F.~Ying.
\newblock Electropaper: Design and fabrication of paper-based electronic
  interfaces for the water environment.
\newblock {\em Electronics}, 10(5), 2021. doi: {{%
10\hspace{.1pt}\discretionary{.}{%
}{.}\hspace{.4pt}3390\discretionary{/}{%
}{/}electronics10050604}}


\bibitem{terre2020focus}
S.~Mareschal and S.~Delaney.
\newblock Using focus group discussions with children and adolescents.
\newblock Terre des Hommes, 2020.

\bibitem{mendez2017bottom}
G.~G. M\'{e}ndez, U.~Hinrichs, and M.~A. Nacenta.
\newblock Bottom-up vs. top-down: Trade-offs in efficiency, understanding,
  freedom and creativity with infovis tools.
\newblock In {\em Proceedings of the 2017 CHI Conference on Human Factors in
  Computing Systems}, CHI '17, p. 841–852. Association for Computing
  Machinery, New York, NY, USA, 2017. doi: {{%
10\hspace{.1pt}\discretionary{.}{%
}{.}\hspace{.4pt}1145\discretionary{/}{%
}{/}3025453\hspace{.1pt}\discretionary{.}{%
}{.}\hspace{.4pt}3025942}}


\bibitem{murphy2013data}
S.~A. Murphy.
\newblock Data visualization and rapid analytics: Applying tableau desktop to
  support library decision-making.
\newblock {\em Journal of Web Librarianship}, 7(4):465--476, 2013.

\bibitem{oblinger2004next}
D.~Oblinger.
\newblock The next generation of educational engagement.
\newblock {\em Journal of interactive media in education}, 2004(1), 2004.

\bibitem{oblinger2005leading}
D.~Oblinger.
\newblock Leading the transition from classrooms to learning spaces.
\newblock {\em Educause quarterly}, 1(7-12), 2005.

\bibitem{papert1993children}
S.~Papert.
\newblock {\em The children's machine: Rethinking school in the age of the
  computer}.
\newblock Basic Books, Inc., 1993.

\bibitem{papert2020mindstorms}
S.~A. Papert.
\newblock {\em Mindstorms: Children, computers, and powerful ideas}.
\newblock Basic books, 2020.

\bibitem{payne2020dance}
W.~C. Payne, Y.~Bergner, M.~E. West, C.~Charp, R.~B.~B. Shapiro, D.~A. Szafir,
  E.~V. Taylor, and K.~DesPortes.
\newblock Danceon: Culturally responsive creative computing.
\newblock In {\em Proceedings of the 2021 CHI Conference on Human Factors in
  Computing Systems}, CHI '21. Association for Computing Machinery, New York,
  NY, USA, 2021. doi: {{%
10\hspace{.1pt}\discretionary{.}{%
}{.}\hspace{.4pt}1145\discretionary{/}{%
}{/}3411764\hspace{.1pt}\discretionary{.}{%
}{.}\hspace{.4pt}3445149}}


\bibitem{perin2021students}
C.~Perin.
\newblock What students learn with personal data physicalization.
\newblock {\em IEEE Computer Graphics and Applications}, 41(6):48--58, 2021.
  doi: {{%
10\hspace{.1pt}\discretionary{.}{%
}{.}\hspace{.4pt}1109\discretionary{/}{%
}{/}MCG\hspace{.1pt}\discretionary{.}{%
}{.}\hspace{.4pt}2021\hspace{.1pt}\discretionary{.}{%
}{.}\hspace{.4pt}3115417}}


\bibitem{punch2002children}
S.~PUNCH.
\newblock Research with children: The same or different from research with
  adults?
\newblock {\em Childhood}, 9(3):321--341, 2002. doi: {{%
10\hspace{.1pt}\discretionary{.}{%
}{.}\hspace{.4pt}1177\discretionary{/}{%
}{/}0907568202009003005}}


\bibitem{resnick2000beyond}
M.~Resnick, R.~Berg, and M.~Eisenberg.
\newblock Beyond black boxes: Bringing transparency and aesthetics back to
  scientific investigation.
\newblock {\em The Journal of the Learning Sciences}, 9(1):7--30, 2000.

\bibitem{resnick2005design}
M.~Resnick, B.~Myers, K.~Nakakoji, B.~Shneiderman, R.~Pausch, T.~Selker, and
  M.~Eisenberg.
\newblock {Design Principles for Tools to Support Creative Thinking}.
\newblock 1 2005. doi: {{%
10\hspace{.1pt}\discretionary{.}{%
}{.}\hspace{.4pt}1184\discretionary{/}{%
}{/}R1\discretionary{/}{%
}{/}6621917\hspace{.1pt}\discretionary{.}{%
}{.}\hspace{.4pt}v1}}


\bibitem{romero2018speeded}
F.~J. Romero-Ramirez, R.~Muñoz-Salinas, and R.~Medina-Carnicer.
\newblock Speeded up detection of squared fiducial markers.
\newblock {\em Image and Vision Computing}, 76:38--47, 2018. doi: {{%
10\hspace{.1pt}\discretionary{.}{%
}{.}\hspace{.4pt}1016\discretionary{/}{%
}{/}j\hspace{.1pt}\discretionary{.}{%
}{.}\hspace{.4pt}imavis\hspace{.1pt}\discretionary{.}{%
}{.}\hspace{.4pt}2018\hspace{.1pt}\discretionary{.}{%
}{.}\hspace{.4pt}05\hspace{.1pt}\discretionary{.}{%
}{.}\hspace{.4pt}004}}


\bibitem{sarama2004technology}
J.~Sarama.
\newblock Technology in early childhood mathematics: Building blocks as an
  innovative technology-based curriculum.
\newblock {\em Engaging young children in mathematics: Standards for early
  childhood mathematics education}, pp. 361--375, 2004.

\bibitem{sarama2004building}
J.~Sarama and D.~H. Clements.
\newblock Building blocks for early childhood mathematics.
\newblock {\em Early Childhood Research Quarterly}, 19(1):181--189, 2004.

\bibitem{Jonathan2020teaching}
J.~Schwabish.
\newblock Teaching data visualization to kids.
\newblock In {\em VisActivities: IEEE VIS Workshop on Data Vis Activities at
  IEEE VIS 2020}, 2020.

\bibitem{shneiderman-covid}
B.~Shneiderman.
\newblock Data visualization's breakthrough moment in the covid-19 crisis.
\newblock {\em Medium}, 2020.

\bibitem{silver2012makey}
J.~Silver, E.~Rosenbaum, and D.~Shaw.
\newblock Makey makey: Improvising tangible and nature-based user interfaces.
\newblock In {\em Proceedings of the Sixth International Conference on
  Tangible, Embedded and Embodied Interaction}, TEI '12, p. 367–370.
  Association for Computing Machinery, New York, NY, USA, 2012. doi: {{%
10\hspace{.1pt}\discretionary{.}{%
}{.}\hspace{.4pt}1145\discretionary{/}{%
}{/}2148131\hspace{.1pt}\discretionary{.}{%
}{.}\hspace{.4pt}2148219}}


\bibitem{thudt2018reflect}
A.~Thudt, U.~Hinrichs, S.~Huron, and S.~Carpendale.
\newblock Self-reflection and personal physicalization construction.
\newblock In {\em Proceedings of the 2018 CHI Conference on Human Factors in
  Computing Systems}, CHI '18, p. 1–13. Association for Computing Machinery,
  New York, NY, USA, 2018. doi: {{%
10\hspace{.1pt}\discretionary{.}{%
}{.}\hspace{.4pt}1145\discretionary{/}{%
}{/}3173574\hspace{.1pt}\discretionary{.}{%
}{.}\hspace{.4pt}3173728}}


\bibitem{informatics7030035}
M.~Trajkova, A.~Alhakamy, F.~Cafaro, S.~Vedak, R.~Mallappa, and S.~R. Kankara.
\newblock Exploring casual covid-19 data visualizations on twitter: Topics and
  challenges.
\newblock {\em Informatics}, 7(3), 2020. doi: {{%
10\hspace{.1pt}\discretionary{.}{%
}{.}\hspace{.4pt}3390\discretionary{/}{%
}{/}informatics7030035}}


\bibitem{vandemoere2010physical}
A.~Vande~Moere and S.~Patel.
\newblock The physical visualization of information: Designing data sculptures
  in an educational context.
\newblock In M.~L. Huang, Q.~V. Nguyen, and K.~Zhang, eds., {\em Visual
  Information Communication}, pp. 1--23. Springer US, Boston, MA, 2010.

\bibitem{verdine2014deconstructing}
B.~N. Verdine, R.~M. Golinkoff, K.~Hirsh-Pasek, N.~S. Newcombe, A.~T.
  Filipowicz, and A.~Chang.
\newblock Deconstructing building blocks: Preschoolers' spatial assembly
  performance relates to early mathematical skills.
\newblock {\em Child development}, 85(3):1062--1076, 2014.

\bibitem{Verhaert2021datablokken}
P.~Verhaert, G.~A. Panagiotidou, and A.~Vande~Moere.
\newblock `datablokken': Stimulating critical data literacy of children through
  the use of a life-size data physicalisation game.
\newblock In {\em VisActivities: IEEE VIS Workshop on Data Vis Activities at
  IEEE VIS 2021}, 2021.

\bibitem{weber1990basic}
R.~P. Weber.
\newblock {\em Basic content analysis}, vol.~49.
\newblock Sage, 1990.

\bibitem{wickham2016constructionism}
C.~Wickham, C.~Girvan, B.~Tangney, A.~Sipitakiat, and N.~Tutiyaphuengprasert.
\newblock Constructionism and microworlds as part of a 21st century learning
  activity to impact student engagement and confidence in physics.
\newblock {\em Constructionism}, pp. 34--43, 2016.

\bibitem{Wilson2002}
M.~Wilson.
\newblock Six views of embodied cognition.
\newblock {\em Psychonomic Bulletin {\&} Review}, 9(4):625--636, Dec 2002. doi:
  {{%
10\hspace{.1pt}\discretionary{.}{%
}{.}\hspace{.4pt}3758\discretionary{/}{%
}{/}BF03196322}}


\bibitem{zheng2020paper}
C.~Zheng, P.~Gyory, and E.~Y.-L. Do.
\newblock {\em Tangible Interfaces with Printed Paper Markers}, p. 909–923.
\newblock Association for Computing Machinery, New York, NY, USA, 2020.

\bibitem{zinovyev2010data}
A.~Zinovyev.
\newblock Data visualization in political and social sciences.
\newblock {\em arXiv preprint arXiv:1008.1188}, 2010.

\end{thebibliography}
